\documentclass[article,12pt]{article} 
\usepackage{apacite}
\usepackage{amscd,amsmath,amssymb,amsfonts,latexsym,mathrsfs,amsthm,mathtools}
\usepackage{graphicx} 
\usepackage{bigints}
\usepackage{setspace} 
\usepackage{graphics,epsfig}
\graphicspath{{figures/}} %Setting the graphicspath
\usepackage{sectsty}
\usepackage{natbib}
\usepackage[english]{babel}
\usepackage{bm}

\usepackage{multirow}
\usepackage{subfigure}
\usepackage[usenames]{color}
\usepackage{rotating}
\usepackage{url}
\usepackage{bbm}
\usepackage{float}
\usepackage{tikz}
\usepackage{pgf}
\usepackage{pdflscape}
\usepackage{xcolor}
\usepackage{colortbl}
\usepackage{bigints}
\usepackage{cancel}

\def\checkmark{\tikz\fill[scale=0.4](0,.35) -- (.25,0) -- (1,.7) -- (.25,.15) -- cycle;}

\newtheorem{rem}{{\bf Remark}}

\usepackage{flushend}
\usepackage{verbatim}
\usepackage{tabularx}
\usepackage{arydshln}
\usepackage{hyperref}
\usepackage[toc,page]{appendix}
\usepackage{pdflscape}

\newcommand{\E}{\mathbb{E}}

\newcommand{\x}{\bm{\theta}}

\newcommand{\y}{{\bf y}}

\newcommand{\post}{\bar{\pi}}

\definecolor{MYCOLOR0}{rgb}{0.92,0.92,0.92}
\definecolor{MYCOLOR}{rgb}{1,1,0}
\definecolor{MYCOLOR2}{rgb}{0.5,1,0.5}
\definecolor{MYCOLOR3}{rgb}{0.88,1,1}

\usepackage{imakeidx}
\makeindex

\title{%Improper priors in astrophysics: a gentle guide \\
On the safe use of prior densities for Bayesian model selection} 
 
\author{F. Llorente$^\dagger$\footnote{Corresponding author: felloren@est-econ.uc3m.es.}, L. Martino$^{\ddagger}$, E. Curbelo$^\dagger$, J. Lopez-Santiago$^\dagger$, D. Delgado$^\dagger$  \\
{\small$^\dagger$  Universidad Carlos III de Madrid,  Legan\'es (Spain).}\\
{\small$^\ddagger$  Universidad Rey Juan Carlos,  Fuenlabrada (Spain).} \\
}

\date{}
 
\begin{document}

\maketitle

\begin{abstract}
The application of Bayesian inference for the purpose of model selection is very popular nowadays. 
In this framework, models are compared through their marginal likelihoods, or their quotients, called Bayes factors. 
However, marginal likelihoods depends on the prior choice. For model selection,  even diffuse priors can be actually very informative, unlike for the parameter estimation problem.  Furthermore, when the prior is improper, the marginal likelihood of the corresponding model is undetermined.
In this work, we discuss the issue of prior sensitivity of the marginal likelihood and its role in model selection.
We also comment on the use of uninformative priors, which are very common choices in practice. Several practical suggestions are discussed and many possible solutions, proposed in the literature, to design objective priors for model selection are described. Some of them also allow the use of improper priors. 
The connection between the marginal likelihood approach and the well-known information criteria is also presented. We describe the main issues and possible solutions by illustrative numerical examples, providing also some related code. One of them involving a real-world application on exoplanet detection.
\newline
\newline
%\citep{Djuric03,Gordon93,Martino15PF}. It has been widely accepted as a valid measure of effective sample size. 
{ \bf Keywords:} 
Model selection, Marginal likelihood, Bayesian evidence, improper priors, information criteria, BIC, AIC, posterior predictive.
\end{abstract}

%\newline
%\newline

\section{Intro}

In the last decades, we observe a growing trend in the use of Bayesian approaches to the problem of inferring the parameters of physical models describing natural processes. Although Bayesian inference has historically been used (e.g. \citep{Robert04,Liu04b}), it is only now becoming more widespread. Nowadays, we can find applications of Bayesian inference methods in fields such as remote sensing \citep{MartinoCompress2,DeepLLORENTE}, astronomy \citep{Feroz2019,Anfinogentov2021}, cosmology \citep{Ashton2021,Ayuso2021}, or optical spectroscopy \citep{Emmert2019,RevModPhys.83.943}. 
\newline
One of the most common problems we may encounter in Bayesian inference is that of model selection. For this purpose, the determination of the Bayes factor is often used. This involves the approximation of the {\it Bayesian evidence}, a.k.a., {\it marginal likelihood},  of the several models. 
The marginal likelihood shows a clear dependence on the choice of the prior probability density functions (pdfs). Many papers propose  diffuse (usually uniform) prior pdfs, in order to avoid biasing the exploration of the parameter space (see, e.g., \citep{Pascoe2020}). In some cases, the selected prior pdfs are  diffuse or even {\it improper} \citep{Gregory2011}. { These ideas have been borrowed from the Bayesian parameter estimation problem, where they are adequate and {\it objective} choices. However, in model selection, the situation is more complex {as} we describe below.}     
\newline
\newline
In a first part of this work, we describe some issues in Bayesian model selection (or hypothesis testing) based on the marginal likelihood computation \citep{llorente2020marginal,chib2001marginal,bos2002comparison}.
First of all, we show how the results can be affected by the choice of the prior. {The typical solution for parameter estimation of using a diffuse prior (which is said {\it uninformative} in this scenario) cannot be considered an objective choice for the marginal likelihood computation. With an objective choice, we refer to a prior selection that attempts to bring impartiality in the model selection problem, and a diffuse prior can be actually a very informative prior for model selection.}
Secondly, this issue becomes even more dramatic when improper priors are employed: the Bayesian parameter estimation with improper priors is allowed if the corresponding posterior is proper, whereas Bayesian model selection with improper priors is {\it not} allowed, due to the fact the marginal likelihood is not completely specified (it is defined up to an arbitrary constant). We describe all these issues by mathematical considerations and several illustrative numerical examples. One of them involves a real-world application for detecting exo-objects (orbiting other stars) based on a radial velocity model.
\newline
 Furthermore, in the second part of this work, we show some possible solutions presented in the literature, such as hierarchical approaches, {\it likelihood-based priors}, and the {\it partial, intrinsic},  {\it fractional} Bayes factors \citep{llorente2020marginal,o1995fractional}, remarking potential benefits and possible drawbacks of each of them. An alternative to the marginal likelihood approach for Bayesian model selection, called {\it posterior predictive} framework \citep[Ch. 6]{vehtari2017practical}\citep{piironen2017comparison}, is also described. Finally, the relationship between the  information criteria \citep{konishi2008information}, such as Bayesian-Schwarz information criterion (BIC),  
Akaike information criterion  (AIC), and  the marginal likelihood approach is discussed in Appendix \ref{ObradeArteCriteria}. Therefore, the  contribution is twofold:   we provide (a) a gentle guide for interested practitioners (with several warnings and advices), and (b) a work useful for more expert researchers looking for practical solutions and/or possible alternatives.  Some related code is also provided.

\section{Problem statement}\label{ModelFitSect}

%{ problema con $\theta_{D_\theta}$ y $\theta_{D_m}$ }
%\subsection{Problem statement}
%%%%%%%%%%%%%%%%%%%%%%%%%%%%%%%%%%%%
In many applications,  the goal is to make inference 
 about a variable of interest,  
$\x=\theta_{1:D_{\x}}=[\theta_1,\theta_2,\ldots,\theta_{D_{\x}}]\in {\bm \Theta}\subseteq\mathbb{R}^{D_{\x}}$,
where $\theta_d\in \mathbb{R}$ for all $d=1,\ldots,D_{\x}$, given a set of observed measurements, ${\bf y}=[y_1,\ldots,y_{D_y}]\in \mathbb{R}^{D_y}$. In the Bayesian framework, one complete model $\mathcal{M}$ is formed by a likelihood function $\ell({\bf y}|\x,\mathcal{M})$ and a prior probability density function (pdf) $g(\x|\mathcal{M})$. All the statistical information is summarized by the posterior pdf, i.e.,
\begin{equation*}
{\bar \pi}(\x|\y,\mathcal{M})= \frac{\ell(\y|\x,\mathcal{M}) g(\x|\mathcal{M})}{p(\y|\mathcal{M})},
\label{eq:posterior}
\end{equation*}
where 
\begin{equation}\label{MarginalLikelihood}
Z= p({\bf y}|\mathcal{M})=\int_{\bm \Theta} \ell({\bf y}|\x,\mathcal{M}) g(\x|\mathcal{M}) d\x,
\end{equation}
is the so-called {\it marginal likelihood}, a.k.a., {\it Bayesian evidence} \citep{Robert04,Liu04b}. 
%{\color{violet} Definir integral $S = \int \ell(\y|\x)d\x$, finito e infinito}
This quantity is important for model selection purposes, as we show below.
However, usually $Z= p({\bf y}|\mathcal{M})$ is unknown and difficult to approximate, so that in many cases we are only able to evaluate the unnormalized target function,
\begin{equation}
\pi(\x|\y,\mathcal{M})=\ell({\bf y}|\x,\mathcal{M}) g(\x|\mathcal{M}) \propto {\bar \pi}(\x|\y,\mathcal{M}).
\label{eq:target}
\end{equation}
{\bf Model Selection and testing hypotheses.} Let us consider now $M$ possible models (or hypotheses), $\mathcal{M}_1,...,\mathcal{M}_M$,  with prior probability mass $p_m =\mathbb{P}\left(\mathcal{M}_m\right)$, $m=1,...,M$. Note that, we can have variables of interest $\x_m=[\theta_{m,1},\theta_{m,2},\ldots,\theta_{m,D_{\x_m}}]\in {\bm \Theta}_m\in \mathbb{R}^{D_{\x_m}}$, with possibly different dimensions in the different models. The posterior probability of the $m$-th  model is given by
\begin{eqnarray*}
p(\mathcal{M}_m|{\bf y})&=&\frac{p_m p({\bf y}|\mathcal{M}_m)}{p({\bf y})}\propto p_mZ_m
\end{eqnarray*}
where $Z_m=p({\bf y}|\mathcal{M}_m)=\int_{{\bm \Theta}_m} \ell({\bf y}|\x_m,\mathcal{M}_m) g(\x_m|\mathcal{M}_m) d\x_m$,
and $p({\bf y})=\sum_{m=1}^M p(\mathcal{M}_m)p({\bf y}|\mathcal{M}_m)$. 
Moreover, the ratio of two marginal likelihoods 
%$\frac{Z_m}{Z_{m'}}$ 
\begin{align*}
\mbox{BF}_{mm'}=\frac{Z_m}{Z_{m'}} =\frac{ p({\bf y}|\mathcal{M}_m)}{p({\bf y}|\mathcal{M}_{m'})} =\frac{p(\mathcal{M}_m|\y)/p_m}{p(\mathcal{M}_{m'}|\y)/p_{m'}},
\end{align*}
also known as {\it Bayes factors}, represents the posterior to prior odds of models $m$ and $m'$.  If some quantity of interest is common to all models, the posterior of this quantity can be studied via {\it model averaging} \citep{BMA99}, i.e., a complete posterior distribution as a mixture of $M$ partial posteriors linearly combined with weights proportionally to $p(\mathcal{M}_m|{\bf y})$  (see, e..g, \citep{Martino15PF,PetarCopy}). 
Therefore, in all these scenarios, we need the computation of $Z_m$ for all $m=1,...,M$. 

\begin{rem}
 Hereafter, {whenever we focus on a single although arbitrary model $\mathcal{M}_m$,} we skip the dependence on $\mathcal{M}_m$ in the notation, for simplicity. For instance, we denote  the posterior density as $\bar{\pi}(\x|\y)$ and  the marginal likelihood as $Z=p(\y)$. Thus, we write %$Z=\int_{\bm \Theta} \pi(\x) d\x$.
\begin{equation}\label{MarginalLikelihood2}
Z=\int_{\bm \Theta} \pi(\x|\y) d\x= \int_{\bm \Theta} \ell(\y|\x) g(\x) d\x.
\end{equation}
\end{rem}

\begin{rem} \label{Rem2}
From Eq. \eqref{MarginalLikelihood2}, we can see clearly that $Z$ is an average of likelihood values $ \ell(\y|\x)$, weighted according to the prior pdf $g(\x)$. 
\end{rem}

\noindent
Clearly, the results of the Bayesian inference depend on the choice of the prior density, the model prior probabilities and the actual number of data $D_y$.
%\newline
%\newline

{
\section{Important definitions and classifications}

In this section, we describe some preliminary definitions that are necessary for a clear description of the issues in Bayesian model selection and the corresponding possible solutions (described in the rest of the work).

\subsection{Levels in Bayesian inference}

%{\color{violet}
%-- estimation and prediction (first level of inference; assume model is right and we aim to estimate/fit parameters)
%
%-- model selection/hypothesis testing (second level of inference, model choice)
%}
%\newline
%\newline
\noindent
Generally speaking, in Bayesian inference we can distinguish between two types of problems or levels of inference \citep[Ch. 28]{mackay2003information}, described below:
\newline
\begin{itemize}
\item {\bf Level-1: estimation and prediction problems.} 
In the first level, given the $m$-th model $\mathcal{M}_m$, we are interested in making inferences regarding parameter $\x_m$ by focusing on its posterior pdf $\post(\x_m|\y,\mathcal{M}_m) \propto \ell(\y|\x_m,\mathcal{M}_m)g(\x_m|\mathcal{M}_m)$. This is also denoted  as ``Level-1 of inference'' in the literature. 
\item {\bf Level-2: model selection problems.} 
In the second type of problem, we focus on the model posterior distribution $p(\mathcal{M}_m|\y) \propto p(\mathcal{M}_m)Z_m = p(\mathcal{M}_m)\int_{{\bm \Theta}_m}\ell(\y|\x_m,\mathcal{M}_m)g(\x_m|\mathcal{M}_m)$ for all $m=1,\dots,M$. This is also known  as ``Level-2 of inference''. 
\end{itemize}
%\newline
%\newline
More {\it levels} of inference can be recognized in the so-called hierarchical  Bayesian approaches.  However, conceptually {these} are the two {\it main} levels of inference since they are associated {with} the two main inference scenarios: parameter estimation and model selection. We will see that the prior choice has a different impact in each of the different levels.  
%\newline

%It is important to remark, the choice of the prior densities  can affect the results of the inference in a different way, depending on whether we are in the first scenario or in the second scenario  \citep{consonni}. A relevant example is the following:  the use (when is possible) of a uniform prior over the entire support ${\bm \Theta}$ is defined as a non-informative prior for parameter estimation, but can be considered {\it highly} informative for the model selection purpose.  We devote the following sections and several part of the rest of the work to clarify this issue. 

}

{
\subsection{Type of model comparison}
}
\noindent
In the literature, we can distinguish different types of model selection,  as we summarize below. The type of model selection problem can affect the user's choice of a suitable prior density. %This is an important  classification since different model selection  
\newline
\begin{itemize}
\item {\bf Basic model selection: } In this scenario, we compare different likelihood functions (i.e., observation models). The likelihood functions can represent completely different models, living even in different parameter spaces. In this scenario, the parameters $\x_m$ of each model can have a completely different physical or statistical interpretation. 
\item {\bf Selection in nested models:} Nested models are models that belong to the same parametric family, but the {\it size} of the model $|{\bm \Theta}_m|=D_{\x_m}$ is also unknown and must be inferred as well, jointly with the parameter $\x_m$. Namely, we have a sequence of likelihoods defined in an increasing dimensional space, such as $\ell({\bf y}|\theta_1,\mathcal{M}_1)$, $\ell({\bf y}|\theta_1,\theta_2,\mathcal{M}_2)$, $\ell({\bf y}|\theta_1,\theta_2,\theta_3,\mathcal{M}_3)$, etc. 
\newline
 Famous applications which belong to this scenario are the following: {\it variable selection} (e.g., selecting a subset of relevant features/variables in regression or classification),  {\it order selection}  (e.g., in polynomial regression or ARMA models etc.), {\it clustering} (when the number of clusters are unknown) and  {\it dimension reduction} problems \citep{bishop2006pattern}. %Note that, in the nested model case, some components of the parameters $\x_m$ are shared in different $\mathcal{M}_m$ (for instance, $\x_2$ and $\x_3$ share the first two components $\theta_1,\theta_2$), and this fact affects the possible prior choice in this framework (see the discussion in Section \ref{ponaqui}).
\end{itemize}
\subsection{Type of prior densities}
%%%%%%%%%%%%%%%%%%

%{\color{violet}
%-- aqui lo de non-informative, default, vague, etc... en fin, todas las priors de forma general y breve $\leftarrow$ definir cada concepto, y agruparlos si es necesario (sin dar consejos)

%-- nuestra definicion de non-informative y discusion para los dos levels of inference (en model selection, no tiene el mismo significado... porque no existe.....)

%-- diferenciar entre uniforme en todo el dominio, y uniforme en un subset (locally uniform parece que se llama...), de forma matematica?

%-- principios de construccion de priors que representan absence of knowdledge... y su relacion con reglas formales

%-- tabla?

%-- 4 cajones: informative-subjective (no datos) sabemos crear la prior (beliefs Jukka, forzar propiedades/regularizers, conjugada); naive non-informative (vague, diffuse, weakly-informative, flat, etc); non-informative objective (obtained by applying formal rules, either ``justified'' or by convention); mezcla...

%}

The literature {has} plenty of works devoted to the specification and classification of different priors. 
The interested readers can find gentle reviews in, e.g.,\citep{kass1996selection,consonni2018prior,mikkola2021prior}.
Here, we provide a brief summary of concepts related to the choice of the  priors $g(\x|\mathcal{M})$ over the parameters, and how this choice can affect the analysis in the two levels of inference {that we have described above.}

{
%%%%%%%%%%%%%%%%%%%
\subsubsection{Subjective priors}
%%%%%%%%%%%%%%%%%%%
If the user or practitioner has some belief or any a-priori knowledge about the quantity of interest (before the data {is} observed), then this information should be included in the analysis by the addition of a suitable prior density. This prior is called as {\it informative} (or more precisely, in our opinion, {\it subjective-informative}).  We can distinguish three main classes of  subjective priors:
\begin{itemize}
\item {\bf Priors including beliefs.} An informative prior pdf can be determined from previous information, past experiments or by other sources of information (different from the observation model).
{Prior elicitation ideas can be used to transform such knowledge into a prior density. See \citep{mikkola2021prior} for a review on different approaches for prior elicitation.}
\item {\bf Priors as regularizers.} In this case, the practitioner/researcher desires to force that the final solution satisfies some properties established in advance, such {as} {\it smoothness} (designing specific structure in covariance matrices in Gaussian priors, e.g., see \citep{martino2021joint}), {\it sparsity} (this is the case of LASSO regularized, i.e., Laplacian priors \citep{bishop2006pattern}) etc. Moreover, the regularization effect produced by the prior usually yields more computational stability, hence reducing the numerical issues.  
\item {\bf Conjugate priors.} A prior can also be set with {the} goal of reducing the computation required by the posterior analysis. Indeed, when a family of conjugate priors exists, choosing a prior from that family simplifies the calculation of the posterior distribution, avoiding the use of costly computational techniques. 
\end{itemize}

%%%%%%%%%%%%%%%%%%%
\subsubsection{Objective priors}
%%%%%%%%%%%%%%%%%%%

In many scenarios, additional information and  conjugate priors are not available, and  ``objective'' choice of priors could be desired \citep{consonni2018prior}.
A first (and perhaps primitive) approach for obtaining an objective prior {is} related to the concept of {\it uninformative} priors,  representing the absence of a-priori knowledge \citep{kass1996selection,consonni2018prior}.
A second approach is related to the idea of constructing priors by the use of formal rules and  automatized procedures based on desirable criteria and properties. The term {\it objective} prior aims at encompassing both groups of priors above \citep{consonni2018prior}.
% Hence, this prior family contains priors that try to represent absence of knowledge and produce minimal impact on the inference, but also priors that are obtained by an automatized procedure (and whose properties are theoretical and/or practically appealing).
 Below, we give more specific definitions.}
\newline
\newline
{\bf Uninformative priors}.  Generally, a prior is defined as {\it uninformative}, if it has been chosen in order to have a minimal impact on the posterior density \citep{mikkola2021prior}. In this sense, for the inference problem of parameter estimation ({\bf Level-1}),   a uniform prior over all the support ${\bm \Theta}$ is the maximal expression of uninformative prior. In fact, the inference would be completely data-driven.   On the contrary,  we will show that in model selection ({\bf Level-2}), this prior is highly informative.  Below, we describe some classes of uninformative priors (or attempts of uninformative priors) for the {\bf Level-1} of inference, i.e., parameter estimation.
\begin{itemize}
 
\item {\bf Uniform prior over ${\bm \Theta}$ when $|{\bm \Theta}|<\infty$.} 
If ${\bm \Theta}$ is bounded, the simplest idea for determining a non-informative prior (for  {\bf Level-1}, parameter estimation)  is to assign equal probabilities to all possible outcomes, such as uniform densities in the bounded support, i.e., $g(\x)\propto 1\ \forall\x\in{\bm \Theta}$. 
\item {\bf Locally-uniform priors.} If  ${\bm \Theta}$ is unbounded, one can employ {\it vague} priors, i.e., densities with probability mass spread in all the state space, with a great scale parameter (this is the reason {for} the name ``locally uniform'').
The priors built using this philosophy   have been given different names such as {\it diffuse}, {\it vague}, {\it  flat}, {\it  weakly-informative}, etc. \citep{consonni2018prior} A more extreme alternative is to use {\it improper} priors when it is possible (see the description below).
\item {\bf Improper priors.} Let us consider again that ${\bm \Theta}$ is unbounded.  The use of improper priors, i.e., such that $\int_{\bm \Theta} g(\x) d\x=\infty$, is allowed for {\bf Level-1} inference when $\int_{\bm \Theta} \ell(\y|\x) g(\x) d\x  <\infty$, since the corresponding posteriors are proper. 
{ The simplest example is {\it the uniform improper prior}, i.e.,  $g(\x)\propto 1$ for all $\x$ in the unbounded support ${\bm \Theta}$. It is often employed for expressing the absence of a-priori information in the {\bf Level-1} of inference. }
%could appear at first glance  as a possible  solution. 
However, improper priors are not allowed for model selection ({\bf Level-2} inference), where we use the marginal likelihood $Z$. Indeed,
the prior $g(\x)=c\cdot h(\x)$ is not completely specified, since $c>0$ is arbitrary. %{\color{red}aqui no faltaria decir algo de improper uniform? si no, no se entiende por que se querrian utilizar como  uninformatives...}
%Some possible solutions are given in Section \ref{AppBayeFactImproperPrior}.  
%A uniform density over an unbounded domain is a clear example of  improper prior. 
%This is a clear example of prior choice that has different interpretations in the two types of inference problems. Indeed, a uniform improper prior can be seen as uninformative for parameter estimation but it does not make sense for model selection.
%Furthermore, proper uniform and locally uniform priors share the same feature of being uninformative for parameter estimation rather than for model selection. 
%In the first illustrative example, we show that an increasingly diffuse prior will inevitably produce a model with null marginal likelihood (namely, the classical Lindley-Bartlett paradox). %{para model selection, prior increasingly diffuse producen mas penalizacion...} 
\end{itemize} 
\noindent
Other authors design priors using formal rules which are theoretical and practically appealing. In this sense, this type of priors are {\it informative but not subjective}. Some example are given below.  
\newline
\newline
{\bf Reference and Jeffreys priors.} Prior densities can also be designed according to other principles such as invariance after transformations, symmetry or maximizing entropy given some constraints \citep{kass1996selection,consonni2018prior}. Examples of this family are the {\it reference} priors \citep{bernardo1979reference,berger2009formal} and {\it Jeffreys} priors \citep{jeffreys1998theory}. Often, they are also improper priors. An example is $g(\sigma) \propto 1/\sigma$ for $\sigma>0$ which is an  improper Jeffreys prior, and is usually applied for a variable that represents a  standard deviation. 
{More generally, the Jeffreys prior is constructed by taking $g(\x) \propto |\mathcal{I}(\x)|^{\frac{1}{2}}$ where $\mathcal{I}(\x)$ denotes the Fisher information matrix.
%The above priors are still popular as uninformative choices in the estimation/prediction problem (See Sect. 2 in \citep{consonni} for a in-depth discussion of these priors). 
}
\newline
\newline
Below we discuss how the choice of the prior affects {\bf (a)} the inference of $\x$ ({\bf Level-1}), and {\bf (b)} the estimation of the Bayesian evidence $Z$ for the model selection problem ({\bf Level-2}).

%%
%
%
%
%
%
%
%
%
%
%
%
%%
%
%
%
%
%%
%
%
%
%
%%
%
%
%
%
%

%%%%%%%%%%%%%%%%%%%%%%%%%%%%%%%
\section{Dependence on the choice of the prior density}\label{SuperSuperIMPSect}
%%%%%%%%%%%%%%%%%%%%%%%%%%%%%%%
%-- comentar algo de objective priors al final...
%\newline
%\newline
%{\bf Benefits.}  

In this section, we show how the marginal likelihood $Z$ depends on the choice of prior density \citep{Bernardo94}.
Here, first we show all the possible values that the evidence $Z$ can take when changing the prior pdf.  Then, we present some reassuring asymptotic results. Finally, we describe further issues with the use of improper priors.
%\newline
%\newline

\subsection{Bounds of the evidence $Z$}\label{sec_bounds_Z}

%{\bf Bounds of the evidence $Z$}. 

Let us denote the maximum and minimum value of the likelihood function as  $\ell_\text{min}=\ell(\y|\x_\text{min})= \min\limits_{\x\in{\bm \Theta}} \ell(\y|\x)$,  and $\ell_\text{max}=\ell(\y|\x_\text{max})=  \max\limits_{\x\in{\bm \Theta}}\ell(\y|\x)$, respectively. Note that
\begin{align*}
	Z = \int_{\bm \Theta} \ell(\y|\x)g(\x)d\x &\leq  \ell(\y|\x_\text{max})\int_{\bm \Theta} g(\x)d\x = \ell(\y|\x_\text{max}).
\end{align*}
Similarly, we can obtain $Z\geq  \ell(\y|\x_\text{min})$. The maximum and minimum value of $Z$ are reached, for instance, with two degenerate choices of the prior, $g(\x) = \delta(\x-\x_\text{max})$ and $g(\x) = \delta(\x-\x_\text{min})$, where $\delta(\x)$ denotes the Dirac point mass at ${\bf 0}$. 
Hence, for every other choice of $g(\x)$, we have 
\begin{align*}
	\ell(\y|\x_\text{min}) \leq Z \leq \ell(\y|\x_\text{max}).
\end{align*}
Namely, depending on the choice of the prior $g(\x)$, we can have any value of Bayesian evidence contained in the interval $[\ell(\y|\x_\text{min}),\ell(\y|\x_\text{max})]$. 
%For further discussion see Section \ref{SectNuovaBella}. 
\newline
The two possible extreme values correspond to the worst and the best model fit, respectively. We can obtain $Z=\ell(\y|\x_\text{min})$  with the choice $g(\x) = \delta(\x-\x_\text{min})$ (which applies the greatest possible penalty to the model), and we obtain $Z=\ell(\y|\x_\text{max})$, with the choice $g(\x) = \delta(\x-\x_\text{max})$ (which does not apply any penalization to the model complexity, i.e., we have the maximum overfitting).  Indeed, $Z=\int_{\bm \Theta} \ell(\y|{\bm \theta}) g({\bm \theta}) d{\bm \theta}$ is by definition  an average of the likelihood values weighted according to the prior. 
\begin{rem}\label{Rem3}
 Depending on  the choice of the prior, the evidence $Z$ can take any possible value in the interval $[\ell(\y|\x_\text{min}),  \ell(\y|\x_\text{max})]$. Hence, in this sense, the prior $g(\x)$ induces a penalization term for the model complexity. { See also Appendix \ref{LucaImplicitPEN} for further details}.
\end{rem}

{
\begin{rem}   Choosing a prior $g(\x)$, we fix our bias-variance trade-off (a point between the maximum under-fitting  and maximum over-fitting). In this sense, the sensitivity of $Z$ could be also considered as a benefit, i.e., an additional degree of freedom for improving the bias-variance trade-off.
\end{rem}
}
\noindent
Note that Remark  \ref{Rem3} above it is strictly connected to Remark \ref{Rem2}.
For the relationship with the well-known Bayesian-Schwarz information criterion (BIC) and the  Akaike information criterion (AIC), see Appendix \ref{ObradeArteCriteria}.
%{
%The marginal likelihood (and more generally, the results of Bayesian inference) depend on both the prior and the number of data.
%}
%\newline
%\newline

\subsection{Asymptotic  considerations in Bayesian inference}

\noindent Throughout this section, we consider the priors have been selected and  fixed, whereas the number of data $D_y$ diverges to infinity, i.e., $D_y \to \infty$.
 We summarize the basic consistency properties of Bayesian inference in both inference problems (i.e., Level-1, estimation, and Level-2, model selection) and discuss the asymptotic behavior of Bayes factors and posterior model probabilities. % e summarize the limiting behavior when $D_y \to \infty$ of the Bayesian posterior distributions, $\post(\x_i|\y,\mathcal{M}_i)$ and $p(\mathcal{M}_i|\y)$, in the two scenarios above.
For sake of simplicity, we assume that weak regularity conditions are satisfied for these results to hold \citep{dawid2011posterior,kass1995bayes,rossell2021balancing,Bernardo94}. 
\newline
 It is important to distinguish and describe two scenarios:  {\bf (a)} one where the true (unknown) distribution of the data $\ell_\text{true}(y|\x_\text{true})$ is included in the $M$ possible models ($\mathcal{M}$-closed scenario), {\bf (b)} and the other one where $\ell_\text{true}(y|\x_\text{true})$ is {\it not} included in the possible set of models ($\mathcal{M}$-open scenario). 
 \begin{itemize}
	\item {\bf $\mathcal{M}$-closed scenario.} When one of the models under consideration, say $\mathcal{M}_{i_\text{true}}$, contains $\ell_\text{true}(y|\x_\text{true})$, i.e., there is $\x_\text{true}$ such that $\ell_\text{true}(y|\x_\text{true}) = \ell(y|\x_\text{true},\mathcal{M}_{i_\text{true}})$.
	
	\item {\bf $\mathcal{M}$-open scenario.} When none of the models contains $\ell_\text{true}(y|\x_\text{true})$ ({\it misspecification}), then we can define  % define for every model $\mathcal{M}_i$,
		$\x_i^* = \arg\min_{\x_i} KL(\ell_\text{true},\ell(\cdot|\x_i,\mathcal{M}_i))$,      
	which is the parameter that minimizes the Kullback-Leibler (KL) divergence between $\ell_\text{true}(y|\x_\text{true})$ and $\ell(y|\x_i,\mathcal{M}_i)$. 
	Furthermore, we can define
	\begin{align*}
		\mathcal{M}_{i^*} = \arg\min_i KL(\ell_\text{true},\ell(\cdot|\x^*_i,\mathcal{M}_i)),
	\end{align*}
	as the model that is closest in KL divergence to the true distribution of the data. 
\end{itemize}
%\newline
%\newline
\noindent{\bf Consistency in Level-1.}
Consider the posterior distribution $\post(\x|\y)$ for a fixed model $\mathcal{M}$ (that is, a particular observation model  and a fixed prior).
In the $\mathcal{M}$-closed scenario, the posterior $\post(\x|\y)$ concentrates around $\x_\text{true}$ as $D_y \to \infty$ (see Bernstein-von Mises theorem \citep{Robert04,Liu04b,Bernardo94}).
Then, the two Bayesian point estimators, the posterior mean
$\widehat{\x}_{\texttt{mean}} =\int_{{\bm \Theta}} \x {\bar \pi}(\x|\y)d\x$, 
and the maximum-a-posteriori (MAP) estimator $
\widehat{\x}_{\texttt{MAP}}=\arg\max_{\x\in {\bm \Theta}} {\bar \pi}(\x|\y)$,
converge to  $\x_{\text{true}}$ (recovering frequentist arguments). 
This means that for large amounts of data, one can use the posterior distribution to make, from a frequentist point of view, valid statements about estimation and uncertainty. 
  { In the $\mathcal{M}$-open scenario (i.e. when the model is misspecified), then the  asymptotic limits of the estimators $\widehat{\x}_{\texttt{mean}}$  and $\widehat{\x}_{\texttt{MAP}}$   approach the best-fitting parameters $\x^*_i$ \citep{Bernardo94,rossell2021balancing}. }
\newline
\newline
{{\bf Consistency in Level-2.}}
In the $\mathcal{M}$-closed scenario, as the sample size diverges, $D_y \to \infty$, the posterior model distribution concentrates around the true model, that is, $p(\mathcal{M}_{i_\text{true}}|\y)\to 1$  \citep{kass1995bayes,dawid2011posterior}.  
In the $\mathcal{M}$--open scenario, the posterior model distribution concentrates on the model closest in KL divergence, that is,  $p(\mathcal{M}_{i^*}|\y) \to 1$, as $D_y \to \infty$ \citep{dawid2011posterior,rossell2021balancing}. 
%The consistency of Bayesian model selection is summarized as follows.  Under very weak conditions \citep{dawid}:
%\begin{itemize}
%	\item When $\{\mathcal{M}_i\}_{i=1}^M$ contains $\mathcal{M}_\text{true}$ (i.e., $\mathcal{M}_{i_*} = \mathcal{M}_\text{true}$ for some $i_*$),  then the posterior probability $p(\mathcal{M}_{i_*}|\y) \to 1$ as the sample size $D_y \to \infty$. 
%	
%	 \item When $\{\mathcal{M}_i\}_{i=1}^M$ does not contain $\mathcal{M}_\text{true}$,  then the posterior probability concentrates on the model that is closest in KL divergence to $\mathcal{M}_\text{true}$, that is, $p(\mathcal{M}_{i_\text{min}}|\y) \to 1$ as the sample size $D_y \to \infty$, where $\mathcal{M}_{i_\text{min}}$ is model that is closest in KL. 
%\end{itemize}
{
\begin{rem}
	Under regularity conditions, Bayesian parameter estimation and model selection are consistent.
	Specifically, as $D_y \to \infty$, in the $\mathcal{M}$-closed scenario, Bayesian inference gives the correct answer by selecting the true model $\mathcal{M}_{i_\text{true}}$, and also converging to $\x_\text{true}$. In the $\mathcal{M}$-open scenario, Bayesian inference gives the best approximate answer, converging to the KL minimizers under each model $\x^*_i$ and selecting the model with overall minimal KL divergence $\mathcal{M}_{i^*}$.
\end{rem}
}

{
\noindent
Furthermore, in specific application frameworks and under fairly general conditions, asymptotic expressions of quotients of posterior model probabilities and Bayes factors have been derived; see,  e.g., \citep{dawid2011posterior,rossell2021balancing}. An important observation is that the leading terms in those expressions do not depend on the prior densities. Namely, in the asymptotic regime, Bayesian model selection is more sensitive to the sample size $D_y$ than to the prior specifications \citep{dawid2011posterior,rossell2021balancing}. 
%{\color{red}Namely, Bayes factors depend more strongly on the sample size $D_y$ than on the prior dispersion, as we show in the example in Section \ref{sec_illus_2}. (esta frase es igual que la anterior....)}
As we can see in Figure \ref{exceptions}(b), there exists  a reasonable ``default range'' of the prior dispersion parameter that provides good results. Such default ranges could be obtained, for instance, by using a measure of predictive accuracy  \citep{rossell2021balancing}.}
\newline
\newline 
These results for the asymptotic regime are reassuring and comforting. 
However, in the finite sample size regime (i.e., $D_y$ fixed) the results of Bayesian model selection are indeed affected by the prior choice: as we already discussed in Sect. \ref{sec_bounds_Z}, the marginal likelihood can take any value in the interval $[\ell_\text{min},\ell_\text{max}]$. Below, we discuss this issue in the context of increasingly diffuse priors, and compare it with Bayesian parameter estimation.

\subsection{Robustness of Bayesian inference to the prior dispersion}

\noindent In this section, we keep the (finite) number of data $D_y$ fixed,  and  we vary the spread of the prior density (changing some hyperparameter of the prior). 
Below, we consider an illustrative example to show the perceived differences in robustness of Bayesian parameter estimation (Level-1) and Bayesian model selection (Level-2). 

%%%%%%%%%%%%%%%%%%%
\subsubsection{Illustrative example}\label{LucaParaSection}
%%%%%%%%%%%%%%%%%%%
Here, we provide an alternative formulation of the Lindley-Bartlett paradox \citep{lindley1957statistical,villa2017mathematics,robert2014jeffreys}) which shows the well-known robustness of the parameter posterior distribution (Level-1)  when increasingly diffuse priors are employed. These priors are common for parameter estimation where they are seen as uninformative.
%Indeed, diffuse priors are seen as informative exactly because the posterior is not changing when we make it more and more diffuse. 
However, in model selection (Level-2), actually such prior{s} are highly informative: an increasingly diffuse prior penalizes more and more the considered model.
\newline
Let us assume a likelihood function that is integrable in every subset of an unbounded ${\bm \Theta}$, that is, for all  $A\subseteq {\bm \Theta}$, $ \int_{A \in \bm{\Theta}} \ell(\y|\x)d\x < \infty$.
In particular, when $A={\bm \Theta}$, the integral corresponds to the ``area below'' the likelihood function 
	\begin{align}
		S =  \int_{{\bm \Theta}} \ell(\y|\x)d\x < \infty.
\end{align}
Hence, in this scenario, the normalized likelihood is a proper pdf on ${\bm \Theta}$. 
Then, we consider a uniform and proper prior defined on the hyper-volume $B$, i.e.,
$$g(\x) = \frac{1}{|B|} \bm{1}_{B}(\x),
$$
where $|B|$ represents the volume of $B$. Hence, the posterior pdf is 
\begin{align}
	\post(\x|\y) = \frac{\ell(\y|\x)\bm{1}_{B}(\x)}{\int_B \ell(\y|\x)d\x},
\end{align}
which is the normalized likelihood restricted to the set $B$.
\newline
\newline
{\it  Level-1 of inference.} As we increase the volume of $B$, more and more mass of the likelihood is considered. Roughly speaking, for a $|B|$ great enough, the posterior is insensitive to further increase the size of $B$. Indeed, as $|B| \rightarrow \infty$, we have that $\post(\x|\y)$ becomes closer and closer to   
\begin{align}
	\post^*(\x|\y) = \frac{\ell(\y|\x)}{\int_{{\bm \Theta}} \ell(\y|\x)d\x}= \frac{\ell(\y|\x)}{S}.
\end{align}
%It is reasonable to expect that, once we reach a size such the important regions of the likelihood (carrying most of the mass) are included within the volume $B$, the posterior is not changing a lot.
Namely,  in the limit where $B={\bm \Theta}$, the prior $g(\x)$ becomes equivalent to an improper uniform prior on $\x$, for which the Bayesian estimators coincide with their frequentist counterparts.  The posterior $\post^*(\x|\y)$ contains only the information {included} in the likelihood function, and is not affected or distorted by the prior. In this sense, when {it} can be used (i.e., $S$ is finite), a uniform improper prior is the maximal expression of a non-informative prior for the Level-1 of inference. 
\newline
\newline
{\it Level-2 of inference.} We focus now on the marginal likelihood $Z$ which, in this case, is given by
	\begin{align}\label{SuperImpZlucaParadox}
		Z = \frac{\int_B \ell(\y|\x)d\x}{|B|}.
%		= \frac{Z_\ell(B)}{V(B)}
	\end{align}
Now, consider increasing $B$ until we cover all parameter space. In this situation,
$$
|B| \to \infty, \quad \text{but} \quad \int_B \ell(\y|\x)d\x \to S,
$$
Hence,
\begin{align}
	\lim_{|B| \to \infty}Z = 0.
\end{align}
We see that the marginal likelihood of a model with a increasingly-diffuse uniform proper prior becomes null.
This is because increasing the spread of the prior penalizes more and more the considered model. Hence, note that,  in Level-2 of inference, a diffuse uniform prior is actually highly informative. 
\newline
\newline
Now, we can already deduce some conclusions, highlighted below.
\newline
\begin{rem}  In the Level-1 of inference, if $S =  \int_{{\bm \Theta}} \ell(\y|\x)d\x$ is finite, we can use a, proper or improper, {uniform} prior as non-informative choice. Moreover, under the assumption of strong data\footnote{{With ``strong data'', we refer to a dataset under which the likelihood function is very concentrated (i.e, many data or data that are very informative).}}, and if we vary the prior density, the estimators $\widehat{\x}_{\texttt{mean}}$, $\widehat{\x}_{\texttt{MAP}}$ do not change drastically. In {this} case, { under mild conditions and by using} an improper uniform prior, we can recover the frequentist results \citep{consonni2018prior}. 
\end{rem}

\begin{rem} {In Level-2 inference, the concept of non-informative prior cannot be applied. Any choice of prior (also a diffuse, flat one) is actually very informative.}   If $S= \int_{{\bm \Theta}} \ell(\y|\x)d\x$ is finite,  diffuse priors tend  to produce smaller values of the marginal likelihood $Z$    \citep{cameron2014recursive,Bernardo94}. Hence, a good model can display a low value of $Z$ only because we choose a prior that is very spread out. Conversely, a worse model can display a bigger value of $Z$ due to choosing a concentrated prior \citep{Bernardo94,mackay2003information,r2019marginal,llorente2020marginal}.
\end{rem}
{
\begin{rem} The evidence $Z$ contains an implicit penalization of the model complexity. See Appendices \ref{LucaImplicitPEN}-\ref{ObradeArteCriteria}  and \citep[Ch. 28]{mackay2003information}\citep{knuth2015bayesian}.
\end{rem}}
%\newline
%\newline

%The speed at which $\log Z$ diverges depends on the relative speed of each term approaching their limit ($\log S$ and $\infty$ respectively) as $\delta\to\infty$.
%We can interpret the first term in the above equation as a fitting term, and the second term as a penalty term depending on the size $\delta$ and the dimensionality $D_{\x}$. 
%\newline 
%This illustrative example  should warn us about the fact that choosing increasingly diffuse priors (which are  motivated from its uninformativeness in parameter estimation) cannot be viewed as uninformative in Bayesian model selection. On the contrary, increasingly diffuse priors apply arbitrarily large penalizations to the models.  
%For this reason, one could obtain any result by diffuse priors to the different models.
%However, as we noted in the previous sections, results of Bayesian model selection are consistent (under fairly general conditions) and are asymptotically more dependent on the sample size. Hence, increasing the sample size in most cases will be able to overcome the indeterminacy of the setting these {\it informative} diffuse priors (connection with Example 2 in SEct. \ref{sec_illus_2})

%and some possible solutions are given in Section \ref{AppBayeFactImproperPrior}. 

\subsection{Issues with improper priors for model selection}\label{sec_issues_improper}

In the previous section, we {just} discussed the sensitivity of $Z$ to variations of the spread of the prior density, and the fact a diffuse prior is highly informative in the Level-2 of inference. Even more caution is needed in the case of employing improper priors. Indeed, we have seen that the use of improper priors, $\int_{\bm \Theta} g(\x) d\x=\infty$, is allowed for Level-1 inference when $\int_{\bm \Theta} \ell(\y|\x) g(\x) d\x  <\infty$, since the corresponding  posteriors are proper. 
%could appear at first glance  as a possible  solution. 
However,  improper priors are not allowed for the Level-2 (model selection).  We describe this fact below and some possible solutions in the rest of the work.
 \newline
The use of improper priors is common in Level-1 of inference to represent weak  a-priori information. Consider $g(\x)\propto h(\x)$ where $h(\x)$ is a non-negative function whose integral over the state space does not converge,  $\int_{\bm \Theta} g(\x)d\x=\int_{\bm \Theta} h(\x)d\x= \infty$. In that case, $g(\x)$ is not completely specified. Indeed, we can have different  definitions $g(\x)=c\cdot h(\x)$ where $c>0$ is (the inverse of) the  ``normalizing'' constant,  not uniquely determinate since $c$ formally does not exist. Regarding the parameter inference and posterior definition, the use of improper priors poses no problems as long as $\int_{\bm \Theta} \ell(\y|\x)h(\x)d\x <\infty$, indeed
\begin{align}
	\post(\x|\y) &= \frac{1}{Z} \pi(\x|\y)=\frac{\ell(\y|\x)ch(\x)}{\int_{\bm \Theta} \ell(\y|\x)ch(\x)d\x} = \frac{\ell(\y|\x)h(\x)}{\int_{\bm \Theta} \ell(\y|\x)h(\x)d\x}, \nonumber \\
	&=\frac{1}{Z_h}\ell(\y|\x)h(\x)
\end{align}
where $Z=\int_{\bm \Theta} \ell(\y|\x)g(\x)d\x$,   $Z_h=\int_{\bm \Theta} \ell(\y|\x)h(\x)d\x$ and $Z=cZ_h$. Note that the unspecified constant $c>0$ is canceled out, so that the posterior $\post(\x|\y)$ is well-defined even with an improper prior if $\int_{\bm \Theta} \ell(\y|\x)h(\x)d\x <\infty$. However, the issue is not solved when we compare different models, since $Z=cZ_h$ depends on {the} undetermined value $c$. For instance, the Bayes factors depend on the undetermined constants $c_1, c_2>0$ \citep{spiegelhalter1982bayes}, 
\begin{align}
	\text{BF}(\y) =\frac{c_1}{c_2} \frac{\int_{\Theta_1} \ell_1(\y|\x)h_1(\x)d\x}{\int_{\Theta_2} \ell_2(\y|\x)h_2(\x)d\x}
	=\frac{Z_{1}}{Z_{2}}=\frac{c_1Z_{h_1}}{c_2Z_{h_2}},
\end{align}
so that different choices of $c_1, c_2$ provide different preferable models.
There exists various approaches for dealing with this issue, as we show in the next section. More generally, we describe different solutions for a safe choice of the priors in the Level-2 of inference.

%\section{Safe scenarios for fair comparisons} \label{sec_safe_use_of_improper_priors}

%%%%%%%%%%%%%%%%%%%%%%%%%%%%%%%%%%%%%%%
\section{{Objective} approaches for Bayesian model selection} %with uninformative  priors}
%%%%%%%%%%%%%%%%%%%%%%%%%%%%%%%%%%%%%%%
In Bayesian inference, the best scenario is surely when the user has strong beliefs that can be translated into informative priors. 
When this additional information is not available, a careful strategy should be employed due to the dependence of the evidence $Z$ with the prior choice $g(\x)$. { Moreover, we have seen that in model selection (Level-2), the concept of non-informative prior cannot be directly applied, since any kind of  prior is actually informative in Level-2. For instance, diffuse priors can be very informative in the Level-2 of inference.}
\newline
We define as a safe scenario, an approach where the choice of the priors is virtually not favoring any of the models (i.e., in some sense, the choice of the priors seeks to obtain {\it impartiality} in the model selection problem \citep{gelman2017beyond}), and the results are not depending on some unspecified constant $c>0$ (as in the case of using improper priors). Below, we describe some scenarios and some possible solutions for reducing, in some way, the dependence of the model comparison on a {\it subjective} choice of the priors.  Many solutions proposed in the literature are data-driven approaches (see Section \ref{aquiVALUCALOCO2}).
In  Section \ref{PostPredAppr},  we also discuss an alternative approach for model selection in Bayesian statistics \citep[Ch. 6]{vehtari2017practical}\citep{piironen2017comparison}.
%\newline
%\newline
%{\bf Same priors.} 
%%%%%%%%%%%%%%%%%%%%%%%
\subsection{Same priors in nested models}
%%%%%%%%%%%%%%%%%%%%%%%

Generally, we are interested in comparing two or more models. The use of the same (even improper) priors is suitable when the models have the same parameters  (and hence also share the same parameter space). %, although some choice can be more favorable for some model,
With this choice,  the resulting comparison seems fair and reasonable. However, this scenario is very restricted in practice.
An exception is when we have nested models, which share some common parameters. As noted in \citep[Sect. 5.3]{kass1995bayes}, in the context of testing hypothesis, many authors consider the use of improper priors for nuisance parameters that appear on both null and alternative hypothesis. Since the nuisance parameters appear on both models, the undetermined multiplicative constants cancel out in the Bayes factor.  
\newline
\newline
%{\bf Hierarchical modelling.} 
%%%%%%%%%%%%%%%%%%%%%
\subsection{Hierarchical modeling}
%%%%%%%%%%%%%%%%%%%%%
Hierarchical models are formed by multiple levels with the purpose of estimating also the {\it hyper}-parameters of the assumed prior densities.  More specifically, additional prior pdfs (called often  {\it hyper}-priors) over the {\it hyper}-parameters of the priors are considered \citep{gelman2013bayesian,Bernardo94}. Below, we provide just a summary of the new terms:   
\begin{itemize}
\item Hyper-parameters: parameters of the prior distributions,
\item Hyper-priors: prior distributions on hyper-parameters.
\end{itemize}
The underlying idea is  to vary the hyper-parameters of the prior pdfs and perform different inference problems. Namely, fixing the hyper-parameters and studying the posterior, we have one inference problem. Then, we change the hyper-parameters and study the corresponding posterior, we have another inference problem.  
Let us consider now that our prior pdf can be expressed as a parametric (or non-parametric) family of functions.
 We can vary the parameters in this family and even make inference on those variables. In this sense, we reduce the dependence on the choice of the prior, since we are not actually considering a unique prior {\it but} a family of them. { For this reason, several authors claim that the resulting (hierarchical) models seem to be more robust than the non-hierarchical versions \citep{Bernardo94}. } 
 \newline 
 Mathematically speaking, let us denote $g(\x|{\bm \nu})$ our family of priors over $\x$ with  hyper-parameters ${\bm \nu}\in \mathbb{R}^\xi$. Below, we discuss two possible solutions. 
 \newline
 {{\bf Empirical Bayes approach.} In this case, we can compute the evidence  in Eq. \eqref{MarginalLikelihood2} as a function of ${\bm \nu}$, i.e.,
  $Z({\bm \nu})=p(\y|{\bm \nu})=\int_{{\bm \Theta}} \ell(\y|\x)g(\x|{\bm \nu})d\x$,
 and then set 
 \begin{equation}\label{EmBayesEq}
{\bm \nu}^*=\arg\max_{{\bm \nu}} Z({\bm \nu}).
\end{equation}
 Thus, we can use $g(\x|{\bm \nu}^*)$ as a prior over the parameter $\x$ in our inferences \citep{liang2008mixtures,petrone2014empirical}. Note that, in this approach, the choice of the prior is in some sense {\it data-driven}, since $ {\bm \nu}^*$ is obtained by the maximization of $p(\y|{\bm \nu})$ (see also Section \ref{aquiVALUCALOCO2}). 
 %Many other data-driven strategies has been suggested in the literature, as we describe  in Section \ref{aquiVALUCALOCO2}. 
 }
  \newline
 {\bf Full Bayesian approach.}  Assuming an hyper-prior $g_h({\bm \nu})$, the complete posterior is given by the following expression,
\begin{equation}
\bar{\pi}(\x,{\bm \nu}|{\bf y}) =\frac{\ell(\y|\x)g(\x|{\bm \nu}) g_h({\bm \nu})}{Z_{\texttt{new}}},
\end{equation}
where 
\begin{align}
Z_{\texttt{new}} =p(\y)&=\int_{\bm \Theta} \int_{\mathbb{R}^\xi} \ell(\y|\x)g(\x|{\bm \nu}) g_h({\bm \nu}) d\x d{\bm \nu},  \\
&= \int_{\mathbb{R}^\xi} Z({\bm \nu}) g_h({\bm \nu}) d{\bm \nu}, 
\end{align}
 is a Bayesian evidence that takes into account all the members of the prior family. %Note that we have set $Z({\bm \nu})=\int_{\bm \Theta}  \ell(\y|\x)g(\x|{\bm \nu})  d\x $
  Clearly, the model selection scheme based on $Z_{\texttt{new}}$ could be consider more robust than a model selection approach based on a single marginal likelihood $Z=Z({\bm \nu})$,  only using one possible value of ${\bm \nu}$ (i.e., only a unique prior). However, the computation of $Z_{\texttt{new}}$ is more complex than the computation of a single $Z({\bm \nu})$, since we have to approximate a  higher dimensional integral \citep{llorente2020marginal}. Also in the empirical Bayes scheme, we need to compute several values $Z({\bm \nu})$'s for different ${\bm \nu}$'s, in order to perform the optimization in \eqref{EmBayesEq}.\footnote{Note that analytical solutions are generally not available.} Hence, this approach can be much more computational demanding.  
  \newline
  Moreover, the hierarchical framework moves  (in some sense) the problem ``to another level'', where we have to choose the hyper-prior $g_h({\bm \nu})$ or,  in the simplest case,  we have at least to decide one possible value ${\bm \nu}^*$ for setting $g(\x|{\bm \nu}^*)$. Even in this last scenario (and when $S$ is finite), we could choose ${\bm \nu}^*$ such that the prior $g(\x|{\bm \nu}^*)$ is diffuse, reducing arbitrarily  the value of the evidence $Z$ (potentially approaching zero). It is also important to notice that this problem is shared with all the modern statistics, machine learning, and signal processing fields.  Indeed, we  always have some parameters to tune that can dramatically change the results (e.g., regularization parameters in Ridge Regression, LASSO, etc. \citep{bishop2006pattern,martino2021joint}). Hence, the real question is whether one can set these tunable parameters to reasonable values.

{
%%%%%%%%%%%%%%%%%%%%%%%%%%%%%%
\subsection{Data-driven and model-based approaches}\label{aquiVALUCALOCO2}
%%%%%%%%%%%%%%%%%%%%%%%%%%%%%%

Here, we describe different strategies for  constructing data-driven or model-based objective priors. Some ideas for using improper priors in the Level-2 of inference, and other possible approaches for Bayesian model selection are also discussed.

\subsubsection{Likelihood-based priors}\label{aquiVALUCALOCO}}

\noindent
In this section, we describe possible simple data-driven ideas for setting the priors, presented in an increasing order of complexity, i.e., starting from the simplest idea and describing  progressively more sophisticated approaches (proposed in the literature). 
%( discussing benefits and drawbacks of each one). More sophisticated approaches have been proposed in the literature, as we described later on. %However, we believe that discussing these simple ideas (with benefits and drawbacks) helps the interested reader. %for a better understanding of the reader.
\newline
\newline
{\bf Idea-1.} When $S=\int_{\bm \Theta} \ell(\y|\x)d\x <\infty$, we can build a proper  prior based on the data and the observation model. For instance, we can choose $g_\text{like}(\x) =\frac{\ell(\y|\x)}{\int_{\bm \Theta} \ell(\y|\x)d\x}$, then the marginal likelihood is 
\begin{align}\label{EqLucaLoco}
	Z = \int_{\bm \Theta} \ell(\y|\x) g_\text{like}(\x)d\x = \frac{\int_{\bm \Theta}  \ell^2(\y|\x)d\x}{\int_{\bm \Theta}  \ell(\y|\x)d\x}. 
\end{align}
We can consider $g_\text{like}(\x)$ a non-subjective prior in the sense that it does not incorporate any additional information, since it is based only on the data. This idea is also connected to the {\it posterior predictive approach}, that is described in Section \ref{PostPredAppr}.  { However, this prior can be very informative and uses the data twice, so other approaches can be designed for dealing with these issues.}
\newline
\newline
%\newline %Indeed, the marginal likelihood above can be written as $Z=E_{{\bar \pi}(\x|\y)}[\ell(\y|\x)]=\int_{\bm \Theta} \ell(\y|\x) {\bar \pi}(\x|\y)d\x$ when $g(\x)=1$. 
{\bf Idea-2.} Less informative likelihood-based priors can be constructed using a tempering effect with a parameter $0<\beta\leq 1$ or considering only a subset of data, denoted as $\y_\text{sub}$. For instance, when $\int_{\bm \Theta} \ell(\y|\x)^{\beta}d\x <\infty$ or $\int_{\bm \Theta} \ell(\y_\text{sub}|\x)d\x <\infty$, we can choose $g_\text{like}(\x) \propto \ell(\y|\x)^{\beta}$ or  $g_\text{like}(\x) \propto \ell(\y_\text{sub}|\x)$, then the marginal likelihood is
\begin{align}\label{eq_Jabali_parcial}
	Z= \frac{\int_{\bm \Theta} \ell(\y|\x)^{\beta+1}d\x}{\int_{\bm \Theta} \ell(\y|\x)^{\beta}d\x}, \quad \mbox{ or }  \quad Z=\frac{\int_{\bm \Theta} \ell(\y|\x) \ell(\y_\text{sub}|\x) d\x}{\int_{\bm \Theta} \ell(\y_\text{sub}|\x) d\x}.
\end{align}
{ However, we still use a subset of the data twice.}
\newline
\newline
{\bf Idea-3: Data partition.}  In order to avoid to use part of the data twice,  we can divide the data in two subsets, $\y=(\y_\text{train},\y_\text{test})$. Then, if  $S_\text{train}=\int_{\bm \Theta} \ell(\y_\text{train}|\x)d\x <\infty$, we use  $g_\text{like}(\x)=\frac{1}{S_\text{train}} \ell(\y_\text{train}|\x)$, obtaining    
\begin{align}\label{eq_Jabali_ATAQUE}
 Z=\frac{\int_{\bm \Theta} \ell(\y_\text{test}|\x) \ell(\y_\text{train}|\x) d\x}{S_\text{train}}.
\end{align}
If the data are conditionally independent given $\x$, we have that $\ell(\y_\text{test}|\x)\ell(\y_\text{train}|\x)=\ell(\y|\x)$  and 
\begin{align}\label{eq_para_partialBF}
	Z=\frac{S}{S_{\text{train}}}.
\end{align} 
A generalization of Eq. \eqref{eq_para_partialBF} can be obtained considering the conditional likelihood $\ell(\y_\text{test}|\x,\y_\text{train})$ such that $\ell(\y_\text{test}|\x,\y_\text{train})\ell(\y_\text{train}|\x) = \ell(\y|\x)$ is always satisfied \citep[Sect. 2]{o1995fractional}.\footnote{Note that we are abusing of the notation by using the same letter ``$\ell$" for different functions, since we have $\ell(\y_\text{test}|\x,\y_\text{train})=p(\y_\text{test}|\x,\y_\text{train})$, whereas $\ell(\y_\text{test}|\x)=p(\y_\text{test}|\x)$ is not conditioned to other data.  } 
In order to build the less possible informative $g_\text{like}(\x)$, we can look for the {\it minimal} training sets $\y_\text{train}=\y_{\text{min}}$, i.e., the sets with a  minimum number of data, such that   $S_{\text{min}}=\int_{\bm \Theta} \ell(\y_{\text{min}}|\x)d\x <\infty$ \citep{berger1996intrinsic}. 
The dependence on the specific partition can be alleviated by averaging over different partitions. Assume that $R$ is the number of considered partitions. Let us also assume that for each possible training set $\y_\text{train}^{(r)}$, we have $S_\text{train}^{(r)}=\int_{\bm \Theta} \ell(\y_\text{train}^{(r)}|\x)d\x <\infty$, for $r=1,...,R$. Thus, we can build $R$ different priors $g_\text{train}^{(r)}(\x)=\frac{1}{S_\text{train}^{(r)}} \ell(\y_\text{train}^{(r)}|\x)$ and then consider a mixture of posterior densities, each one with a different prior $g_\text{train}^{(r)}(\x)$.  
 In this case,  we obtain  $Z=\frac{1}{R}\sum_{r=1}^R\frac{S}{S_\text{train}^{(r)}}$, where recall that $S=\int_{\bm \Theta} \ell(\y|\x)d\x$. 
  This approach  is related to the  {\it partial} and {\it intrinsic} Bayes factors \citep{o1995fractional,berger1996intrinsic}.
\newline
\newline
{\bf Connection with partial and intrinsic Bayes factors.} 
Let $g_\text{base}(\x)$ denote an improper baseline prior. 
We already discussed that using improper priors produces marginal likelihoods that are specified up to an arbitrary constant (see Sect. \ref{sec_issues_improper}).
Partial Bayes factors (PBFs) are solutions proposed for dealing with this issue, and are based on the same idea of training the prior using some partial likelihood \citep[Sect. 2]{o1995fractional}.
As a result, each model is assigned a marginal likelihood in the form of Eq.  \eqref{eq_para_partialBF}, but also considering the improper baseline  $g_\text{base}(\x)$, i.e., 
\begin{align}\label{eq_Z_IBF}
	Z = \frac{\widetilde{Z}}{\widetilde{Z}_\text{train}} = \dfrac{\int \ell(\y|\x)g_\text{base}(\x)d\x}{\int \ell(\y_\text{train}|\x)g_\text{base}(\x)d\x}.
\end{align}
%where $\widetilde{Z} = \int \ell(\y|\x)g_\text{base}(\x)d\x$ and $\widetilde{Z}_\text{train} = \int \ell(\y_\text{train}|\x)g_\text{base}(\x)d\x$.
Note that any arbitrary constant contained in $g_\text{base}(\x)$ is canceled out in the computation of $Z$. 
Hence, the final Bayes factor (called partial Bayes factor) between any two models is
\begin{align}\label{BFBayeslaw}
	\text{BF}_{12}(\y_\text{test}|\y_\text{train}) = \frac{Z_1}{Z_2} = \frac{\widetilde{Z}_1/\widetilde{Z}_{\text{train},1}}{\widetilde{Z}_2/\widetilde{Z}_{\text{train},2}} = \frac{\widetilde{Z}_1/\widetilde{Z}_2}{\widetilde{Z}_{\text{train},1}/\widetilde{Z}_{\text{train},2}} = \frac{\text{BF}_{12}(\y)}{\text{BF}_{12}(\y_\text{train})},
\end{align}
where we have denoted $\text{BF}_{12}(\y)=\frac{\widetilde{Z}_1}{\widetilde{Z}_2}$ and $\text{BF}_{12}(\y_\text{train})=\frac{\widetilde{Z}_{\text{train},1}}{\widetilde{Z}_{\text{train},2}}$.  Clearly, we should take $\y_\text{train}$ of minimal size.
As above, in order to reduce the sensitivity of the results, we can average $\text{BF}_{12}(\y^{(r)}_\text{test}|\y^{(r)}_\text{train})$ over the possible $R$ partitions, leading to the intrinsic Bayes factors \citep{berger1996intrinsic}. 
%The idea of dividing the data and use some part to ``train'' the prior is also considereded in the ``Partial Bayes Factors'' \citep[Sect. 2]{o1995fractional}.
%Here, the motivation is to be able to compute meaningful marginal likelihoods under improper priors.
%
%
%As a solution for this, we can assign each model with a marginal likelihood in the form of Eq. \eqref{eq_para_partialBF} (where an improper baseline prior is considered, see below.... si no se habla de improper baseline no se entiende)}.
% 
% }
 \newline
\newline
%{
{\bf {\bf Idea-4}: Powered likelihood.} Another alternative given in the literature is the following.  We can use a powered likelihood $\ell(\y|\x)^\beta$ with $0<\beta <1$ to obtain the prior, and employ as likelihood also a tempered version,  i.e., $\ell(\y|\x)^{1-\beta}$, so that we have
\begin{align*}
	g_\text{like}(\x)=g(\x|\beta) \propto \ell(\y|\x)^{\beta}, \quad \mbox{and} \quad  \post(\x|\y) \propto \ell(\y|\x)^{1-\beta}\ell(\y|\x)^\beta,
\end{align*}
Note that, in this case, we do not need the conditionally independent assumption to express the marginal likelihood as ratio of normalizing constants, i.e., 
\begin{align}\label{eq_Z_powered_like}
Z &= \int \ell(\y|\x)^{1-\beta}	g(\x|\beta) d\x =\frac{\int \ell(\y|\x)^{1-\beta}\ell(\y|\x)^\beta d\x}{\int \ell(\y|\x)^\beta d\x} = \frac{S}{S_\beta}. 
\end{align}
	Furthermore, we get rid of the indeterminacy of choosing the partition. However, a tempering value $\beta\in(0,1)$ must be selected. This idea is also employed in the so-called {\it fractional} Bayes factors \citep{o1995fractional}.
\newline
\newline
{\bf Connection with fractional Bayes factors.}
Fractional Bayes factors (FBFs) are another strategy proposed for dealing with an improper baseline $g_\text{base}(\x)$.
This time each model is assigned a marginal likelihood analogous to that of Eq. \eqref{eq_Z_FBF} but considering the baseline prior $g_\text{base}(\x)$, i.e.
\begin{align}\label{eq_Z_FBF}
	Z = \frac{\widetilde{Z}}{\widetilde{Z}_\beta} = \dfrac{\int \ell(\y|\x)g_\text{base}(\x)d\x}{\int \ell(\y|\x)^\beta g_\text{base}(\x)d\x}.
\end{align}
This marginal likelihood is free of arbitrary constants.
The final Bayes factor (called {\it fractional} Bayes factor) between any two models is given as
\begin{align*}
	\text{FBF}_{12} = \frac{Z_1}{Z_2} = \dfrac{\widetilde{Z}_1/\widetilde{Z}_{\beta,1}}{\widetilde{Z}_2/\widetilde{Z}_{\beta,2}} = \frac{\widetilde{Z}_1/\widetilde{Z}_2}{\widetilde{Z}_{\beta,1}/\widetilde{Z}_{\beta,2}} =\frac{\text{BF}_{12}(\y)}{\text{BF}_{12}(\y|\beta)},
\end{align*}
where we denoted $\text{BF}_{12}(\y|\beta) = \frac{\widetilde{Z}_{\beta,1}}{\widetilde{Z}_{\beta,2}}$. Note that FBFs uses again the idea of transforming an improper baseline $g_\text{base}(\x)$ into a proper posterior by conditioning on a tempered likelihood $\ell(\y|\x)^{\beta}$. 
 \newline
 \newline
 {
{\bf Idea-5:  Power-prior.} 
In the literature, other approaches with simulated data have been proposed  \citep{consonni2018prior}. 
Let $\y^*$ denote some imaginary data (i.e., artificial/simulated data) and consider the following {\it power-prior} \citep{ibrahim2015power}
\begin{align}\label{eq_priorSimData}
	g_\text{like}(\x)=g(\x|\y^*,\beta) \propto
	\ell(\y^*|\x)^\beta g_{\text{base}}(\x), \mbox{ where } \quad 0<\beta<1.
\end{align}
 %If we consider $\y^*$ are fixed, we can directly study the marginal likelihood of the posterior $\post(\x|\y,\y^*,\beta) \propto \ell(\y|\x) g(\x|\y^*,\beta)$, similarly to the other ideas above.
%This approach is clearly connected to the likelihood-based priors and the partial, intrinsic, fractional Bayes factors.
An important special case of power priors is the well-known {\it g-prior}, which is an standard prior choice in linear models \citep{zellner1986assessing,liang2008mixtures}. %Moreover, any mixture of g-priors can be considered a power prior with hyper-prior on $\beta$. 
A mixture of g-priors is an objective choice designed for the linear regression setting, that fulfills desirable model selection criteria \citep{bayarri2012criteria}.
\newline
Two further generalizations have been proposed in the literature.
If we consider $\y^*$ are not fixed, but random, we can take an additional step consisting in averaging the prior in Eq. \eqref{eq_priorSimData} with respect to the distribution of the simulated data $\y^*$. 
The resulting prior is thus
\begin{align*}\label{eq_exp_prior}
	g_\text{like}(\x)=g(\x|\beta) = \int g(\x|\y^*,\beta) q(\y^*)d\y^*,
\end{align*}
where $q(\y^*)$ is the distribution of the artificial data. 
With $\beta=1$, the above expression is called {\it expected posterior prior} (EPPs) \citep{perez2002expected}.
Moreover, in the case where all likelihoods (including that of the posterior) are raised to a common power $\beta$ and normalized, we obtain the so-called {\it power expected posterior prior} (PEP priors) \citep{fouskakis2015power}. }
\newline
\newline
Note that most of the approaches described above require $S =  \int_{{\bm \Theta}} \ell(\y|\x)d\x$ be finite, otherwise they cannot be applied. However, in this case, the problem is extended to the Level-1 of Bayesian inference since the posterior would be not proper using a uniform improper prior. 
%{\color{red}esta ultima frase hay que corregirla... When $S=\infty$, the choice of a uniform improper prior could not be possible even in the inference Level 1}

%All these ideas above are also the key idea underlying other techniques below. %the partial and intrinsic Bayes factors described in the next section.

%{\color{violet}hablar de un minimal $\y_\text{sub}$ con Z finita, y hablar de promediar, .... engancha con los fractional BFs de abajo}

 {
 %%%%%%%%%%%%%%%%%%%%%%%%%%%%%%%%%%%%
 \subsubsection{Other model-based approaches for building the prior}
%%%%%%%%%%%%%%%%%%%%%%%%%%%%%%%%%%%%

Other {relevant} ways of designing objective priors consider the information contained in the Fisher information matrix, 
\begin{align}
	\mathcal{I}(\x) = \E_{\ell(\y|\x)}\left[\left(\frac{\partial}{\partial\x}\log \ell(\y|\x)\right)^2\right],
\end{align}
where the expectation is w.r.t. $\ell(\y|\x)$ (fixing $\x$). With the Jeffreys approach,  one takes the prior to be $g(\x) \propto [\mathcal{I}(\x)]^{-\frac{1}{2}}$. This prior has the property of being invariant under change of variables \citep{kass1996selection}.
\newline
The {\it unit information prior} (UIP) is based on the idea that the information encoded in a prior pdf should be roughly the amount of information contained in a single data \citep{consonni2018prior}.
The Fisher information matrix divided by the number of data, i.e., $\frac{1}{D_y}\mathcal{I}(\bm{\mu})$, is thus proposed as an estimate of this information. For instance, for a continuous parameter, $\x\in\mathbb{R}^{d_{\x}}$, we can take the following Gaussian prior,
$$
g(\x) = \mathcal{N}\left(\x\Big| \bm{\mu},\left[\frac{1}{D_y}\mathcal{I}(\bm{\mu})\right]^{-1}\right),
$$ 
where $\bm{\mu}$ is a prior mean. 
%This strategy has important con
In linear models, the UIP takes the same form as the g-prior \citep{consonni2018prior}.
Furthermore, the use of UIP is motivated since it produces a log-Bayes factor that is asymptotically equivalent to the BIC \citep{kass1996selection,consonni2018prior}. 

}

\subsubsection{Posterior predictive approach} \label{PostPredAppr}

 %Other ways of model selection: the posterior predictive approach
 
%{\color{violet}
%-- Poner ref de que este approach no es consistente....
%
%-- ref a los posterior BFs
%
%-- 
%} 

\noindent The marginal likelihood approach is not the only option for model selection in Bayesian statistics. We discuss an alternative strategy, called {\it predictive model selection}, that is based on the concept of prediction  \citep[Ch. 6]{vehtari2017practical}\citep{vehtari2012survey,piironen2017comparison}. This approach is more robust with respect to the choice of the prior density, so it can be considered as a possible solution to the issues described above.
 \newline 
After fitting a Bayesian model, a popular approach for model checking (i.e. assessing the adequacy of the model fit to the data) consists in measuring its predictive accuracy \citep{vehtari2017practical,piironen2017comparison}.  Hence, a key quantity in these approaches is the posterior predictive distribution of generic different data $\widetilde{{\bf y}}$ given ${\bf y}$,
\begin{align}\label{aquiPPA}
p(\widetilde{{\bf y}} | {\bf y})
=E_{\bar{\pi}({\bm \theta} |{\bf y})}[\ell(\widetilde{{\bf y}} | \boldsymbol{\theta})]&=\int_{\bm \Theta} \ell(\widetilde{{\bf y}} | \boldsymbol{\theta}) {\bar \pi}(\boldsymbol{\theta} | {\bf y}) d \boldsymbol{\theta}, \nonumber\\
&=\frac{1}{Z}\int_{\bm \Theta} \ell(\widetilde{{\bf y}} | \boldsymbol{\theta})  \ell({\bf y} | \boldsymbol{\theta})  g({\bm \theta})  d \boldsymbol{\theta},
\end{align}
Considering $\widetilde{{\bf y}}={\bf y}$, we can observe that it exists a clear connection with likelihood-based priors described in Section \ref{aquiVALUCALOCO}. Indeed, if we assume $g(\x)\propto 1$ and $\widetilde{{\bf y}}={\bf y}$, Eq. \eqref{aquiPPA} becomes Eq. \eqref{EqLucaLoco}.
\newline
Note that the posterior predictive distribution in Eq. \eqref{aquiPPA} is an expectation w.r.t. the posterior, which is robust to the prior selection with informative data, unlike the marginal likelihood as we showed in Section \ref{SuperSuperIMPSect}.
{
With a generic $g(\x)$ and $\widetilde{\y}=\y$, the above expression can {be} seen as a marginal likelihood obtained using the posterior as a prior pdf, stressing even more
% {the connection with} 
the approach in Idea-1 described in Section \ref{aquiVALUCALOCO}. It can be also considered as a ``posterior'' Bayes factor, in the sense that the likelihood is averaged w.r.t. the posterior, rather than the prior \citep{aitkin1991posterior}.} 
{
In \citep{djuric1994model}, the predictive density in Eq. \eqref{aquiPPA} is employed to derive predictive Bayesian model selection criteria in the context of normal linear regression with multiple data sequences. This paper explores how one should combine the different predictive densities resulting from the different partitions into training and validation. 
}
Clearly, {these strategies are} less affected by the {initial} prior choice.
\newline
Note that we can consider posterior predictive distributions $p(\tilde{{\bf y}} | {\bf y})$ for vectors $\tilde{{\bf y}}$ smaller than ${\bf y}$ (i.e., with less components). The posterior predictive checking is based on the main idea of considering simulated data $\widetilde{{\bf y}}_{i} \sim p(\widetilde{{\bf y}} | {\bf y})$, with $i=1, \ldots, L$, and comparing {them} with the observed data ${\bf y}$. After obtaining a set of fake data $\left\{\tilde{{\bf y}}_{i}\right\}_{i=1}^{L}$, we have to measure the discrepancy between the true observed data ${\bf y}$ and the set $\left\{\widetilde{{\bf y}}_{i}\right\}_{i=1}^{L} .$ This comparison can be made with test quantities and graphical checks (e.g., posterior predictive p-values) \citep{vehtari2017practical}. A drawback of predictive model selection is that consistency (i.e., selecting the true model as $D_y \to \infty$) is not generally ensured \citep{vehtari2012survey}.
%\newline
{	
	{\rem 	Using the marginal likelihood in Eq. \eqref{eq_Jabali_ATAQUE} or \eqref{eq_Z_IBF} (i.e. Idea-3 and PBFs) as a model selection criterion amounts to selecting the model with greater predictive accuracy. 
	In fact, they are predictive densities of data $\y_\text{test}$ conditional on $\y_\text{train}$ \citep{djuric1990predictive,djuric1994model}. See also Table \ref{table_fer_loco}.}
}

%\begin{center}
	\begin{table}
		{
		\caption{\label{table_fer_loco} Connection between the likelihood-based solutions of Sect. \ref{aquiVALUCALOCO}, with the predictive approach in Eq. \eqref{aquiPPA}.}
		\vspace{-0.3cm}
		\begin{center}
			\begin{tabular}{||c | c | c | c || c | c||} 
		\cline{1-4} 
		\multicolumn{4}{||c||}{Elements in Eq. \eqref{aquiPPA}} & \multicolumn{2}{c}{ }	 \\
				\hline
				$p(\widetilde{\y}|\y)$ & $\ell(\widetilde{\y}|\x)$ & $\ell(\y|\x)$ & $g(\x)$ & Approach & Uses data twice\\ [0.5ex] 
				\hline\hline
				Eq. \eqref{EqLucaLoco} & $\ell(\y|\x)$ & $\ell(\y|\x)$ & 1 & Idea-1 & \checkmark\\ 
				%		\hline
				Eq. \eqref{eq_Jabali_parcial} & $\ell(\y|\x)$ & $\ell(\y_\text{sub}|\x)$ & 1 & Idea-2 & \checkmark \\
				Eq. \eqref{eq_Jabali_parcial} & $\ell(\y|\x)$ & $\ell(\y|\x)^\beta$ & 1 & Idea-2 & \checkmark \\
				Eq. \eqref{eq_Jabali_ATAQUE} & $\ell(\y_\text{test}|\x)$ & $\ell(\y_\text{train}|\x)$ & 1 & Idea-3 &   \\
				%		\hline
				Eq. \eqref{eq_Z_IBF} & $\ell(\y_\text{test}|\x)$ & $\ell(\y_\text{train}|\x)$ & $g_\text{base}(\x)$ & PBFs & \\
				%		\hline
				Eq. \eqref{eq_Z_powered_like} & $\ell(\y|\x)^{1-\beta}$ & $\ell(\y|\x)^\beta$ & 1 & Idea-4 & \\ 
				Eq. \eqref{eq_Z_FBF} &  $\ell(\y|\x)^{1-\beta}$ & $\ell(\y|\x)^\beta$ & $g_\text{base}(\x)$ & FBFs &  \\[1ex] 
				\hline
			\end{tabular}
		\end{center}	
}
	\end{table}
%	\end{center}

%%%%%%%%%%%%%%%%%%%%%%%%%%%%%%%%%%%%%%%%%%
\section{Numerical experiments}\label{NumExp}
%%%%%%%%%%%%%%%%%%%%%%%%%%%%%%%%%%%%%%%%%%

In this section, we provide different numerical simulations testing different models, prior pdfs and possible solutions. One of them is a well-known model based on the radial velocity technique for detecting exo-objects orbiting other stars \citep{Gregory2011,Barros2016}. Some related code is also provided.\footnote{Related Matlab code is available at \url{http://www.lucamartino.altervista.org/Code_Llorente_Priors.m}}

%%%%%%%%%%%%%%%%%%%%%%%55
%%%%%%%%%%%%%%%%%%%%%%%%5
\subsection{Experiment 1}\label{sec_illus_1}

\begin{figure}[h!]
	\centering
	\centerline{
		\subfigure[$\sigma_0=3$]{
			\includegraphics[width=0.4\textwidth]{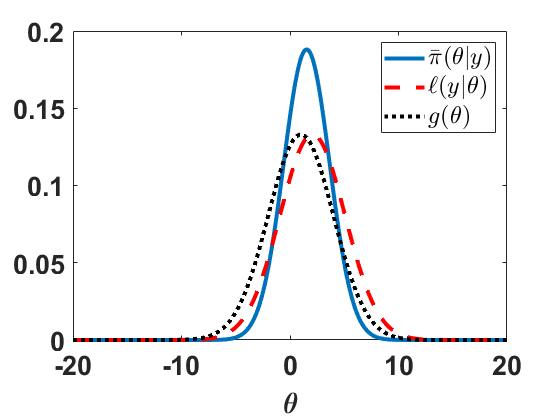}	
		}
		\subfigure[$\sigma_0=10$]{
			\includegraphics[width=0.4\textwidth]{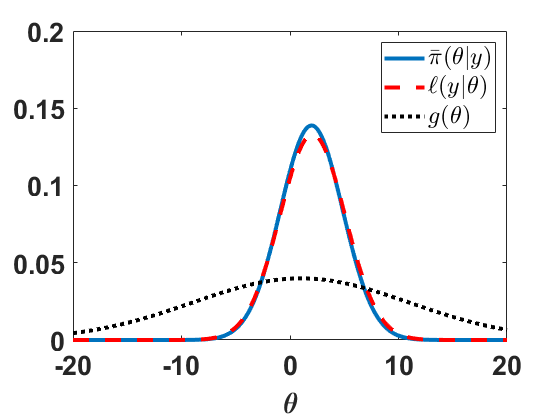}	
		}
	}
	\centerline{
		\subfigure[$\sigma_0=100$]{
			\includegraphics[width=0.4\textwidth]{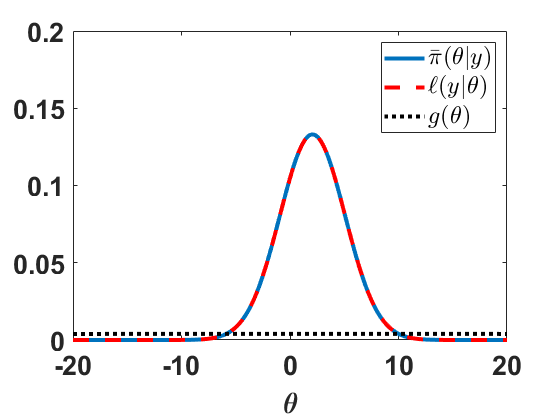}	
		}
		\subfigure[Marginal likelihood $Z$ versus $\sigma_0$]{
			\includegraphics[width=0.4\textwidth]{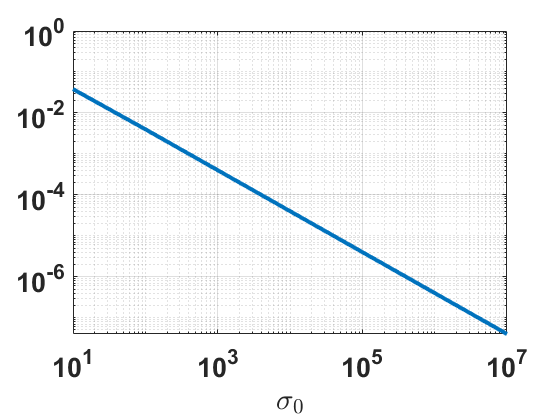}	
		}
	}
	\caption{
		In {\bf (a)}-{\bf (c)}, we show the posterior for a Gaussian prior $\mathcal{N}(\mu_0,\sigma_0^2)$ with three different choices of $\sigma_0$.
		In {\bf (d)}, we show the corresponding marginal likelihood versus $\sigma_0$ in log-scale.
		Note that increasing $\sigma_0$ (i.e. prior is more diffuse) does not change the shape of the posterior, but the marginal likelihood is indeed decreasing.
	}
	\label{Z vs s0}
\end{figure}

%The next illustrative example shows the robustness of the posterior (and the corresponding estimators) and the sensitivity of the evidence $Z$, under prior changes. 
 Let us consider the following Gaussian conjugate model for $\theta$,
\begin{align*}
	\ell(\y|\theta) &= \mathcal{N}(\y|\theta,\sigma^2) = \prod_{i=1}^{D_y}\mathcal{N}(y_i|\theta,\sigma^2) \\
	g(\theta) &= \mathcal{N}(\theta|\mu_0,\sigma_0^2).
\end{align*}
Hence, the posterior is also Gaussian, $\post(\theta|{\bf y})=\mathcal{N}(\theta|\mu_\text{post},\sigma_{\text{post}}^2)$, where
\begin{align*}
	\mu_\text{post} &= \frac{1}{\frac{1}{\sigma_0^2}+\frac{D_y}{\sigma^2}}\left(\frac{\mu_0}{\sigma_0^2}+\frac{D_y\bar{y}}{\sigma^2}\right)\\
	\sigma_\text{post}^2 &= \left(\frac{1}{\sigma_0^2}+\frac{D_y}{\sigma^2}\right)^{-1},
\end{align*}
where $\bar{y}$ denotes the sample mean of ${\bf y}$. 
The marginal likelihood is given by
$$
Z = (2\pi D_y \sigma_n^2)^{-\frac{D_y}{2}}
\left(\frac{\sigma_0^2}{\sigma_n^2}+1\right)^{-\frac{1}{2}}
\exp\left(
-\frac{1}{2}\left(
\frac{v_y+\bar{y}^2}{\sigma_n^2} + \frac{\mu_0^2}{\sigma_0^2} - \frac{1}{\frac{1}{\sigma_n^2}+\frac{1}{\sigma_0^2}}\left(\frac{\bar{y}}{\sigma_n^2}+\frac{\mu_0}{\sigma_0^2}\right)^2
\right)
\right),
$$
where $\sigma_n = \frac{\sigma}{\sqrt{D_y}}$ and $v_y$ denotes the sample variance of ${\bf y}$.
We consider a single data point ($D_y=1$), where ${\bf y}=y=2.078$. 
We fix $\mu_0$ and vary $\sigma_0$.
In Figure \ref{Z vs s0}, we show the corresponding posterior for $\sigma_0=3,10,100$ in solid line, whereas the likelihood is depicted with dashed line and the prior is shown with dotted line. 
The evolution of the corresponding marginal likelihood $Z$ versus $\sigma_0$  is given in Figure \ref{Z vs s0}(d).
%\newline
\newline
As $\sigma_0$ grows, the posterior pdf approaches the likelihood as depicted in Figures \ref{Z vs s0}(a)-(b)-(c). Then, for large values of $\sigma_0$, the posterior is insensitive to {further} increasing the prior dispersion. If we consider $\sigma_0\rightarrow \infty$ (corresponding to an {\it improper} prior),  the posterior pdf coincides with the likelihood function, and the inference (e.g., the  estimators $\widehat{{\bm \theta}}_{\texttt{MMSE}}$ and $\widehat{{\bm \theta}}_{\texttt{MAP}}$) is completely driven by the observed data. In this example  both  estimators $\widehat{{\bm \theta}}_{\texttt{MMSE}}$ and $\widehat{{\bm \theta}}_{\texttt{MAP}}$ converges to the maximum of the likelihood function as  $\sigma_0\rightarrow \infty$. Note also from Figure \ref{Z vs s0}(a) to Figure \ref{Z vs s0}(c) that the variation of the posterior is also negligible. Hence, the improper {uniform} prior is non-informative for Level-1 of inference.
On the contrary, as $\sigma_0$ grows, the marginal likelihood decreases approaching zero as shown in Figure \ref{Z vs s0}(d) (instead of converging to the normalizing constant of the likelihood, as someone could expect). {
This result is consequence of the Jeffrey-Lindley-Bartlett paradox \citep{lindley1957statistical,villa2017mathematics}}. This shows that diffuse priors are very informative in Level-2 of Bayesian inference.
%Based on this fact, many authors claim that the use of Bayes factor is not reliable. 
%\newline
%However, the problem is not that different priors can lead to smaller or bigger values of $Z$ under the same model, the ``better'' prior being always a delta centered in the MLE (as discussed in the toy example in \citep{r2019marginal}). The problem is that over-diffuse priors can make good models have a very low value of $Z$ as compared to other not-so-good models with less diffuse priors. 
%Next, we provide another illustrative example showing that using an increasingly diffuse prior makes us eventually select the wrong model over the true one. \\

%%%%%%%%%%%%%%%%%%%%%%%%%%%%%%%%%%%%%%%%%%%%%%%%%%%%%%%%%%%

\subsection{Experiment 2: Normal linear regression}
%{\bf Normal linear regression example.} 

\begin{figure}[h!]
	\centering
	\centerline{
		\subfigure[$BF_{01}$]{
			\includegraphics[width=0.4\textwidth]{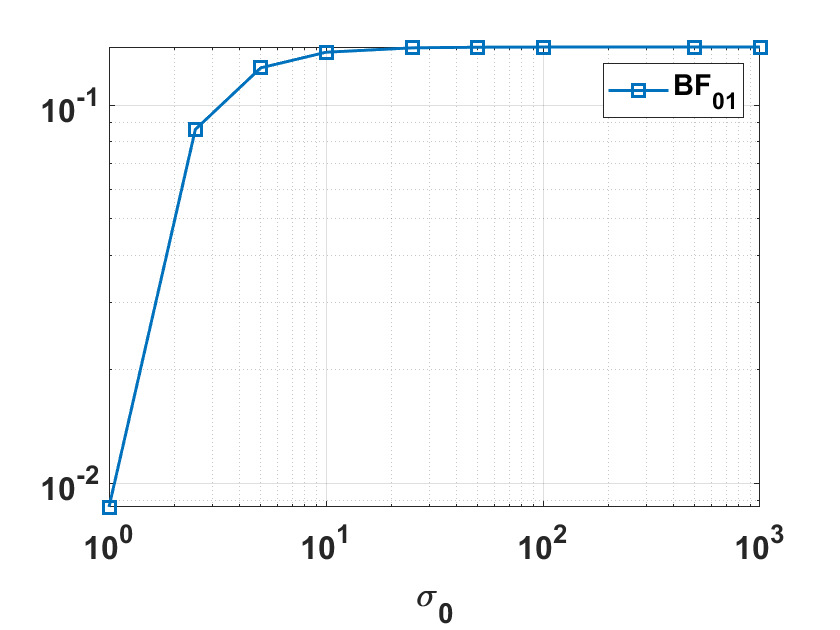}	
		}
		\subfigure[$Z_0$ and $Z_1$]{
			\includegraphics[width=0.4\textwidth]{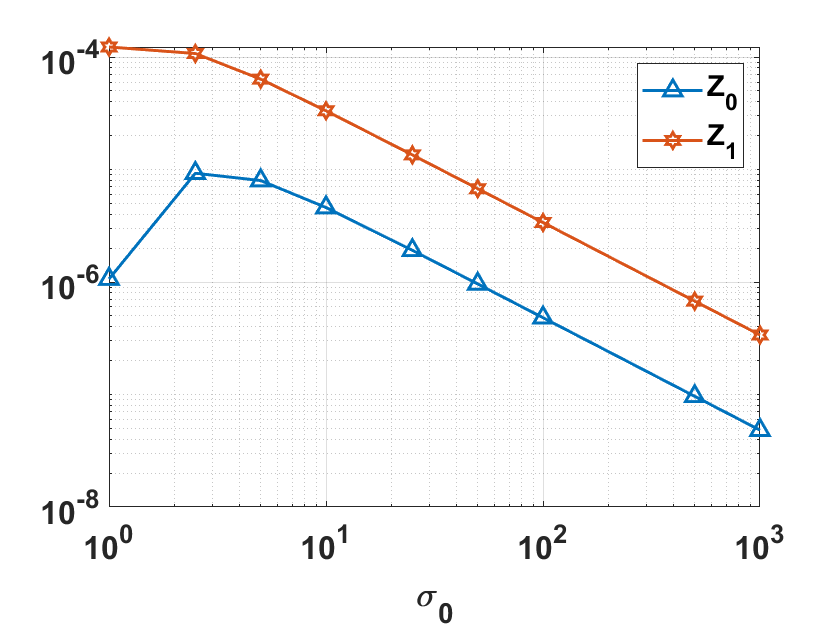}	
		}
	}
	\caption{In {\bf (a)} Bayes factor versus $\sigma_0$ in log-scale. In {\bf (b)} Marginal likelihoods of models $\mathcal{M}_0$ and $\mathcal{M}_1$ versus $\sigma_0$ in log-scale. We consider $\sigma_1 = 1$ is fixed.
	}
	\label{fig_nueva_barriendo_sig0}
\end{figure}

Let us consider the normal linear regression setting with two models for the observations $\y = \{y_i\}_{i=1}^{D_y}$,
\begin{align*}
	\mathcal{M}_0:&\ y_i = \beta_0 + \epsilon_i,   \\
	\mathcal{M}_1:&\ y_i = \beta_0 + \beta_1x_i + \epsilon_i,
\end{align*}
where ${\bf x} = \{x_i\}_{i=1}^{D_y}$ are fixed/known and $\epsilon_i \sim \mathcal{N}(0,\sigma_{\text{like}}^2)$ with $\sigma_{\text{like}}$ known. 
Hence, model $\mathcal{M}_0$ has parameter $\x_0 = \beta_0$, and model $\mathcal{M}_1$ has parameter $\x_1 = [\beta_0,\beta_1]^\top$. 
We set Gaussian priors for both models,
\begin{align}
	g_0(\beta_0) &= \mathcal{N}(\beta_0|0,\sigma_0^2)\quad \text{and} \quad
	g_{1}(\beta_0,\beta_1) = g_0(\beta_0)\mathcal{N}(\beta_1|0,\sigma_1^2).
\end{align}
We aim to analyze the sensitivity of the Bayes factor BF$_{01}$, given by
\begin{align}
	\text{BF}_{01} = \frac{Z_0}{Z_1} = \dfrac{\int \ell(\y|\beta_0)g_0(\beta_0)d\beta_0}{\int \ell(\y|{\bf x},\beta_0,\beta_1)g_1(\beta_0,\beta_1)d\beta_0d\beta_1},
\end{align}
%where $g_{1}(\beta_0,\beta_1) =g_0(\beta_0)\mathcal{N}(\beta_1|0,\sigma_1^2)$, 
when we vary different features such as the dispersions $\sigma_0$ and $\sigma_1$.
\newline
\newline                                                          
\noindent {\bf Sensitivity  w.r.t. the choice of $\sigma_0$.} 
We generate $D_y=4$ observations from model $\mathcal{M}_1$ with $\beta_0^\text{true} = \beta_1^\text{true} = 1$.
We consider $\sigma_1 = 1$ fixed and compute BF$_{01}$ for a sequence of increasing values of $\sigma_0$. 
The Bayes factor BF$_{01}$ versus $\sigma_0$ is shown in Figure \ref{fig_nueva_barriendo_sig0}(a).
It can be seen that BF$_{01}$ is much lower than 1 for every $\sigma_0$, indicating that $\mathcal{M}_1$ is the preferred model.
As expected, BF$_{01}$ is stable under increasing $\sigma_0$, reaching a plateau at $\sigma_0=10$ and becoming constant from there on.
This is a well-known fact: the choice of prior for the common parameter $\beta_0$ does not affect much the comparison.
This is a consequence of choosing the same prior $g_0(\beta_0)$ for both models. In Figure \ref{fig_nueva_barriendo_sig0}(b),  we see that increasing $\sigma_0$ reduces the marginal likelihood of both models simultaneously, hence the pitfalls of using increasingly diffuse priors are solved when we compute the quotient.

\begin{figure}[h!]
	\centering
	\centerline{
		\subfigure[$BF_{01}$]{
			\includegraphics[width=0.4\textwidth]{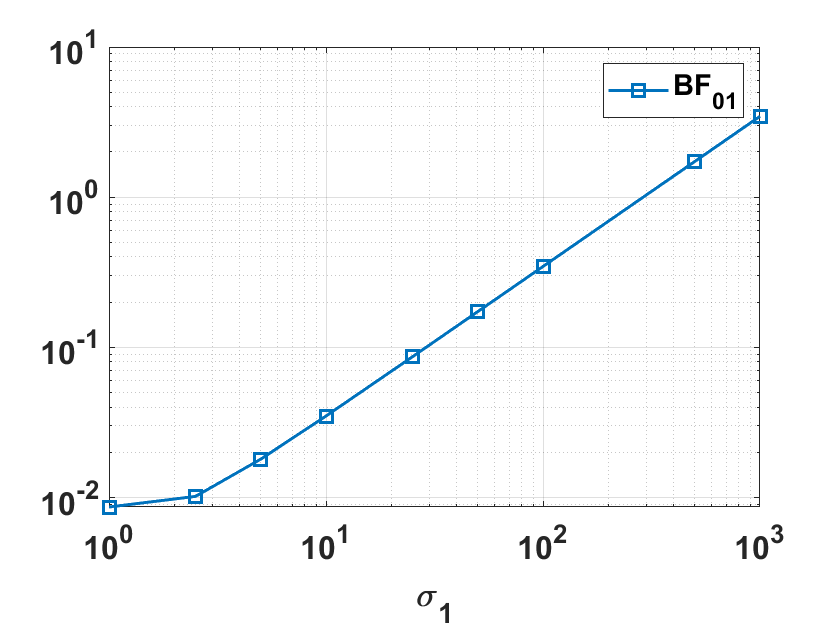}	
		}
		\subfigure[$Z_0$ and $Z_1$]{
			\includegraphics[width=0.4\textwidth]{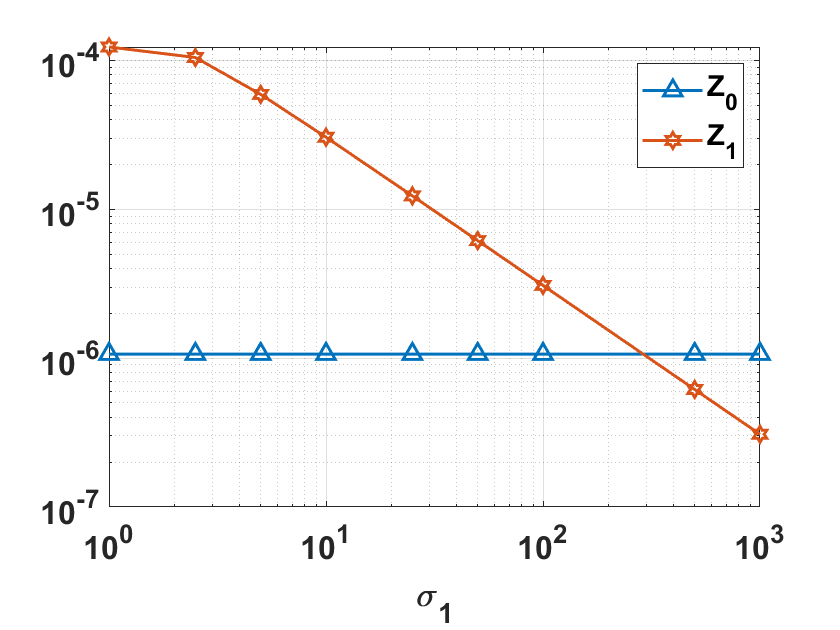}	
		}
	}
	\caption{In {\bf (a)} Bayes factor versus $\sigma_1$ in log-scale. In {\bf (b)} Marginal likelihoods of models $\mathcal{M}_0$ and $\mathcal{M}_1$ versus $\sigma_1$ in log-scale. We consider $\sigma_0 = 1$ is fixed.
	}
	\label{fig_nueva_barriendo_sig1}
\end{figure}

\noindent {\bf Sensitivity w.r.t. the choice of $\sigma_1$.} 
We repeat the experiment but considering a fixed $\sigma_0 = 1$, and compute BF$_{01}$ for a sequence of increasing values of $\sigma_1$. 
The Bayes factor BF$_{01}$ and both marginal likelihoods $Z_0, Z_1$ versus $\sigma_1$ are shown in Figure \ref{fig_nueva_barriendo_sig1}.
Opposite to the previous case, this time we see that BF$_{01}$ is greater than 1 when $\sigma_1 > 500$. 
Indeed, in Figure \ref{fig_nueva_barriendo_sig1}(b), we see that only $Z_1$ decreases as $\sigma_1$ {increases}. 
This is {because} we are only varying the dispersion of the prior in model $\mathcal{M}_1$ not $\mathcal{M}_0$. 
As a consequence, increasing the dispersion of the prior on $\beta_1$ makes us eventually choose the wrong model $\mathcal{M}_0$ (again, this is the Lindley-Bartlett paradox).

\begin{figure}[h!]
	\centering
	\centerline{
		\subfigure[Prior-expected $R^2$]{
			\includegraphics[width=0.4\textwidth]{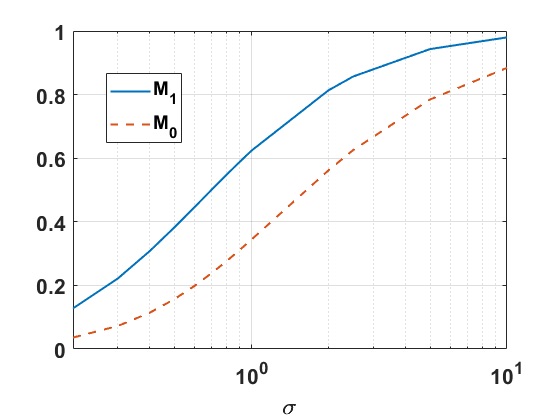}	
		}
		\subfigure[$BF_{01}$]{
			\includegraphics[width=0.4\textwidth]{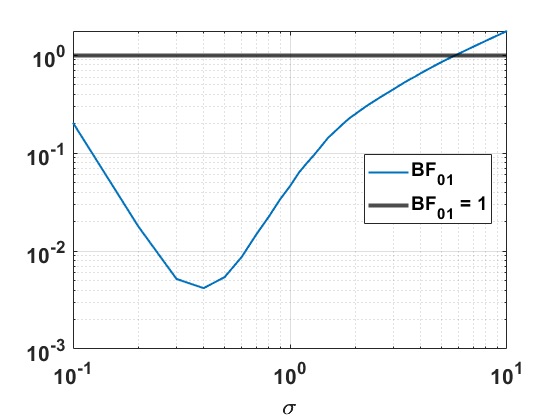}	
		}
	}
	\caption{{{\bf (a)} Prior expected $R^2$ coefficient of models $M_0$ and $M_1$ as a function of $\sigma$. In {\bf (b)} Bayes factor versus $\sigma$ in log-scale.}
	}
	\label{fig_ExpR2teo_and_BF}
\end{figure}

{
\noindent {\bf Sensitivity w.r.t. the choice of $\sigma_0=\sigma_1=\sigma$.} The choice of the prior dispersion can be guided attending to the a-priori predictive power of the model \citep{rossell2021balancing}.  Let $\sigma^2=\sigma_0^2=\sigma_1^2$ denote the (diagonal) variance of the Gaussian priors associated to models $M_0$ ($\x_0 = \beta_0$) and $M_1$ ($\x_1 = [\beta_0,  \beta_1]^\top$), i.e., 
$$
g_0(\x_0) = \mathcal{N}(\beta_0|0,\sigma^2), \quad g_1(\x_1) = \mathcal{N}(\x_1|{\bf 0},\sigma^2{\bf I}_2).
$$
In linear regression, we can observe
%, as a function of the prior dispersion $\sigma$, 
the prior-expected contribution to the signal-to-noise ratio of each model,
$$
\E_{g_m(\x_m)}[w(\x_m)] = \E_{g_m(\x_m)}\left[\frac{\x_m^\top{\bf X}_m^\top{\bf X}_m\x_m}{D_y\sigma^2_{\text{like}}}\right],
$$
where ${\bf X}_m$ denotes the $D_y \times D_{\x_m}$ design matrix of model $m$, or the prior-expected $R^2$ coefficient
$$
\E_{g_m(\x_m)}[R^2(\x_m)] = \E_{g_m(\x_m)}\left[\left(1 + \frac{1}{w(\x_m)}\right)^{-1}\right].
$$
The values of $\E[w(\x_m)]$ or $\E[R^2(\x_m)]$ can help us decide the prior dispersion, which is  modified by the choice of the standard deviation $\sigma$. For instance, the unit information prior (UIP) is obtained by setting the prior dispersion of the model such $\E[w(\x_m)]$ equals the number of parameters \citep{rossell2021balancing}. 
Moreover, there is a range of prior dispersions that produce reasonable values of $\E[w(\x_m)]$ or $\E[R^2(\x_m)]$.
Figure \ref{fig_ExpR2teo_and_BF}(a) shows that for values of $\sigma$ within $[0.1,10]$, the models $M_0$ and $M_1$ display values of $\E[R^2(\x_m)]$, from close to null predictive power, $\E[R^2(\x_m)] = 0$, to perfect predictive power, $\E[R^2(\x_m)]=1$.
Hence, considering for $\sigma$ values only inside this range of values is well justified. 
\newline
Figure \ref{fig_ExpR2teo_and_BF}(b) shows the $BF_{01}$ versus $\sigma$ within $[0.1,10]$ (averaged over repeated independent simulations). We observe that, in this experiment, the Bayesian model selection approach provides always the correct result, when the value of $\sigma$ is selected within the range of {\it reasonable values} discussed above.
}

%We can show specific examples for the setting we are discussing....

%
%
%
% 

%%%%%%%%%%%%%%%%%5
%%%%%%%%%%%%%%%%%
\subsection{Experiment 3}
\label{sec_illus_2}

%{\bf Illustrative example 2.} 

%{\color{violet} remarcar comparacion entre sensibilidad a la dispersion y a los datos; hablar de la consistencia}

\subsubsection{First analysis}

\begin{figure}[!h]
	\centering

	\centerline{
	\subfigure[]{
		\includegraphics[width=0.5\textwidth]{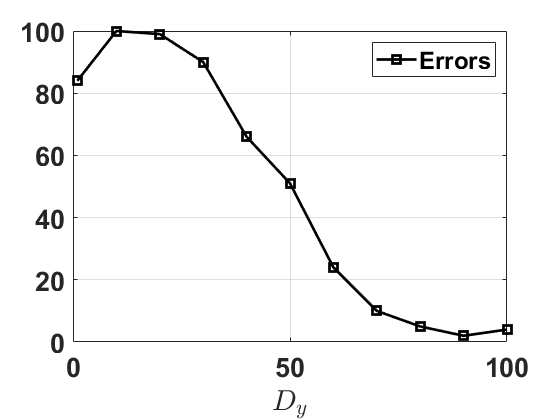}	
	}	
	\subfigure[]{
		\includegraphics[width=0.5\textwidth]{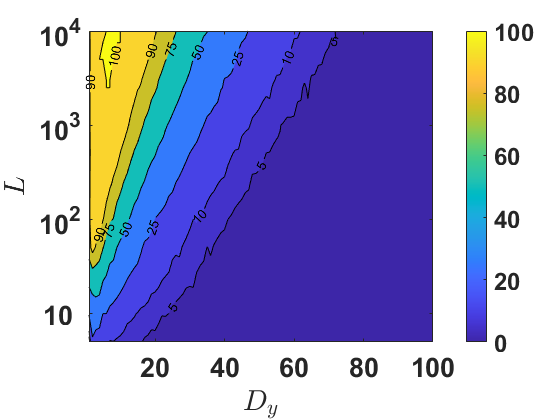}	
	}	
}
	
	\caption{\small In {\bf (a)} number of errors in model selection, i.e., selecting the wrong model ($BF_{12}<1$), out of 100 independent runs, when using $g_1(\theta)=\frac{1}{L},\ \theta\in(0,L)$ with $L=10^5$ (i.e., fixing the prior), for different number of data $D_y$. 
	We can see that, keeping fixed the priors, as $D_y$ grows we choose the true model. However, with fixed $D_y$, changing $L$ we can always adulterate the result of the study penalizing more and the model 1, as shown in Tables \ref{tabla_fer}-\ref{tabla_fer2}.
	{ {\bf (b)} The average number of errors for different values of $L$ and $D_y$.}}

	\label{exceptions}	
	
\end{figure}

Let us consider the problem of selecting between two models,  $\mathcal{M}_1 = \{\ell_1(y|\theta) = \theta^ye^{-\theta}/y!,\  g_1(\theta)\}$ and $\mathcal{M}_2 = \{\ell_2(y|\phi) = \phi(1-\phi)^y,\ g_2(\phi)\}$, namely a Poisson and a geometric distribution \citep{lindley1957statistical}. 
We use a uniform prior $g_2(\phi)=1$ for the proportion $\phi\in[0,1]$, and also a uniform prior $g_1(\theta)=\frac{1}{L}$ for $\theta \in [0,L]$.
We generate $D_y$ independent data ${\bf y} = (y_1,\dots,y_{D_y})$ from $\mathcal{M}_1$ with $\theta_\text{true}=2$.  
The goal of this  example is to show empirically the sensitivity of the Bayes factor to increasing $L$ (i.e., $g_1(\theta)$ becomes more diffuse), and the number of data $D_y$. 
For doing this, for each pair of values $(L, D_y)$, we study the average number of {\it errors} in model selection (i.e., the number of times BF$_{12}<1$) in 100 independent simulated datasets of size $D_y$.
%we repeat this process in  $100$ different runs, and count {\it the number of errors} in model selection, i.e., whenever $BF_{12}<1$ (note it should be greater than 1 since $\mathcal{M}_1$ is the true model).
%\begin{align}
%	BF_{12} = \frac{Z_1}{Z_2}=\frac{\int_0^L \ell_1({\bf y}|\theta)\frac{1}{L}d\theta}{\int_0^1\ell_2({\bf y}|\phi)d\phi}
%\end{align}
%We fix $L$, generate $D_y$ data from model $\mathcal{M}_1$ and compute $BF_{12}$. Hence, the model $\mathcal{M}_1$ is the {\it correct} one, since we generate the data according to $\mathcal{M}_1$.
\newline
First, we compute the number of errors as we increase $L$ for two fixed sample sizes, $D_y = 30$ and $D_y = 100$.
Table \ref{tabla_fer} shows the results when $D_y=30$ for the values $L=10^\alpha$ ($\alpha=1,\dots,6$).
Specifically, we show the maximum and minimum values of BF$_{12}$, obtained in the 100 simulations, along with the number of errors. 
As expected, as $L$ increases, i.e., we use a more diffuse prior, the model $\mathcal{M}_2$ is (wrongly) selected more often. In fact, with $L=10^6$, the Bayes factor always selects $\mathcal{M}_2$ over $\mathcal{M}_1$ (i.e., the Lindley-Bartlett paradox).
Table \ref{tabla_fer2} shows results when $D_y=100$.
On the contrary, we observe here that the number of errors is very low even for large $L$, namely, having more data compensates the potential drawbacks of using a very diffuse prior. 
In addition, in Figure \ref{exceptions}(a), we have computed the number of errors (over the $100$ different runs) for fixed $L=10^5$ versus the number of data $D_y$. 
We see that, for a given prior width, increasing $D_y$ rapidly reduces the number of times we choose the wrong model. 
{Figure \ref{exceptions}(b) shows the average number of errors as a function of both $L$ and $D_y$. We can see again that for fixed $L$, the number of errors is very sensitive to increasing $D_y$.} Namely, a small increase in sample size produces a large reduction in the average number of errors (i.e. the results are consistent).
On the other hand, the number of errors is rather insensitive  to increasing $L$, as compared to $D_y$.
In fact, for $D_y>50$, the number of errors remains constant and close to 0 for all the considered values of $L$ (up to $L=10^4$).
Although increasing $L$ eventually gives the wrong results, this effect is noticeable only when the sample size is small enough.
% This example shows that even if  the number data grows  $D_y$,  diffuse priors can affect the results. 
\newline
Clearly, keeping fixed the (proper) priors, and including the {\it enough} number  of data $D_y$ in our study, we can obtain the correct results (see Figure  \ref{exceptions}). However,  the number of {\it enough} data is unknown and depends on the specific problem. Furthermore, the joint use of a huge amount of data often jeopardized the performance of the computational methods employed for estimating the evidence $Z$ \citep{llorente2020marginal,bos2002comparison}.  
%{In this toy example, }

\begin{table*}[!ht]
	
	\centering
	%\small
	\caption{ Model comparison for $D_y=30$. Minimum and maximum $BF_{12}$ under true model $\mathcal{M}_1$ (Poisson) for 100 simulations.}
	\vspace{0.2cm}
	%	{\footnotesize
	\begin{tabular}{|c|c|c|c|}
		\hline	
		% & & & & & & &  \\
		\multicolumn{4}{|c|}{True model $=\mathcal{M}_1$ (with $\theta_{\text{true}}=2$)}  \\		  
		\hline
		$L$ & min & max & Errors in model choice, over 100 simulations \\
		\hline 
		10 &  0.094 & 4.77$\times 10^{5}$ & 3  \\ 
		\hline
		$10^2$ & 0.059  & 2.49$\times 10^{4}$ & 15 \\
		\hline
		$10^3$& 0.0012 & 1.46$\times 10^{3}$ & 31 \\
		\hline
		$10^4$&1.06$\times 10^{-4}$ & 339.86&67 \\
		\hline
		$10^5$& 1.02$\times 10^{-4}$ & 41.05 & 84 \\
		\hline
		$10^6$& 1.59$\times 10^{-6}$ & 0.7080 & 100 \\
		\hline
	\end{tabular}
	%	}	
	\label{tabla_fer}	
	
\end{table*}
\begin{table*}[!ht]
	
	\centering
	%\small
	\caption{ Model comparison for $D_y=100$. Minimum and maximum $BF_{12}$ under true model $\mathcal{M}_1$ (Poisson) for 100 simulations.}
	\vspace{0.2cm}
	%	{\footnotesize
	\begin{tabular}{|c|c|c|c|}
		\hline	
		% & & & & & & &  \\
		\multicolumn{4}{|c||}{True model $=\mathcal{M}_1$ ($\theta_{\text{true}}=2$)}  \\		  
		\hline
		$L$ & min & max & Errors in model choice, over 100 simulations \\
		\hline 
		10 & 41.27  & 9.05$\times 10^{13}$ & 0   \\ 
		\hline
		$10^2$ & 6.93  & 1.55$\times 10^{13}$ & 0 \\
		\hline
		$10^3$& 14.45 & 2.21$\times 10^{11}$ & 0 \\
		\hline
		$10^4$& 7.94$\times 10^{-4}$ & 3.75$\times 10^{11}$ &  3 \\
		\hline
		$10^5$& 0.5214 & 1.36$\times 10^{12}$  &  2\\
		\hline
		$10^6$& 7.98$\times 10^{-4}$ & 2.07$\times 10^{8}$ & 7 \\
		\hline
	\end{tabular}
	%	}	
	\label{tabla_fer2}	
	
\end{table*}

%%%%%%%%%%%%%%%%%%%%%%%%%%%%%%%%%%%%%%%%%%%%%%%%

%{\bf Illustrative example 3.} 
\subsubsection{Using partial and intrinsic BFs}
\label{sec_intrinsicBF}

%Consider again the comparison between a Poisson distribution and a geometric distribution.
Previously, we considered two uniform and proper priors $g_2(\phi)=1,\ \phi\in(0,1)$, and  $g_1(\theta) = \frac{1}{L},\ \theta\in (0,L)$. Hence, the Bayes factor is well defined. Here, we replace $g_1(\theta)$ with an {\it improper} uniform prior $\widetilde{g}_1(\theta) \propto 1$, $\theta \in (0,\infty)$ for model $\mathcal{M}_1$.  Our goal is to replicate Tables \ref{tabla_fer} and \ref{tabla_fer2} using this  improper prior for $\mathcal{M}_1$.
\newline
In this situation, the Bayes factor is not well-defined due to the arbitrary constant in $\widetilde{g}_1(\theta)$. Hence, we need to resort to partial Bayes factors (PBFs) \citep[Sect. 2]{o1995fractional}, where we compute the posterior of a single observation $y_i$, denoted by a sub-index $i$, (training set) under prior $\widetilde{g}_1(\theta)$, i.e.,  $\post_{1}(\theta|y_i) \propto \ell_1(y_i|\theta)\widetilde{g}_1(\theta)$,
and use $\post_{1}(\theta|y_1)$ now as a proper prior in the computation of BF$_{12}$.
%We consider a single observation as the training sample, ${\bf x}_\text{train} = x_1$. 
In order to avoid the dependence on the training sample, we use the intrinsic Bayes factor (IBF) approach \citep{berger1996intrinsic}. Let  ${\bf y}_{-i}$ denote the vector of all $D_y$ data without the $i$-th  component $y_i$, i.e., ${\bf y}_{-i}$ is a vector of $D_y-1$ components. The IBF consists in averaging over all possible training samples, resulting in 
\begin{align}
\text{IBF}_{12} = \frac{1}{D_y}\sum_{i=1}^{D_y}\frac{\int_0^\infty \ell_1({\bf y}_{-i}|\theta)\post_1(\theta|y_i)d\theta}{\int_0^1\ell_2({\bf y}|\phi)d\phi} = 
\frac{1}{D_y}\sum_{i=1}^{D_y}\frac{\int_0^\infty \ell_1({\bf y}|\theta)d\theta/\int_0^\infty \ell_1( y_{i}|\theta)d\theta}{\int_0^1\ell_2({\bf y}|\phi)d\phi}.
\end{align}
Note that the cost of computing IBF$_{12}$ increases with $D_y$. For this experiment, we generate data from both models with different values of $\theta_\text{true}$ and $\phi_\text{true}$, that is, we alternatively consider $\mathcal{M}_1$ and $\mathcal{M}_2$ as the true model.
We compute IBF$_{12}$ in $100$ different runs  for the chosen values 
of $\theta_\text{true}$ and $\phi_\text{true}$, and we show the results in 
Table \ref{table_1} and Table \ref{table_2} for $D_y=30$ and $D_y=100$, respectively\footnote{Related Matlab code is available at \url{http://www.lucamartino.altervista.org/Code_Llorente_Priors.m}}.
We show the maximum and minimum values of IBF$_{12}$, obtained in the 100 simulations, along with the number of errors.
When $\mathcal{M}_1$ is the true model, IBF$_{12}<1$ corresponds to an error, and conversely, when $\mathcal{M}_2$ is the true model, IBF$_{12}>1$ corresponds to an error. 
\newline
The results clearly show that the use of intrinsic Bayes factors allows for correctly selecting $\mathcal{M}_1$ when it is indeed the true model, with very few errors in model selection for the considered values of $\theta_\text{true}$ and both $D_y=30$ and $D_y=100$. On the contrary, when $\mathcal{M}_2$ is the true model, the use of intrinsic Bayes factors  makes more probable selecting $\mathcal{M}_1$ for some values of $\phi_\text{true}$. Note, for instance, that the number of errors when $\phi_\text{true}=0.8$ is 66, that is, more than half of the times we would wrongly select $\mathcal{M}_1$ over $\mathcal{M}_2$.
This is consistent with the idea underlying PBF and IBF, where the proper prior is built using part of the data. Indeed, it tends to artificially increase the marginal likelihood of the model where the likelihood-based prior is applied (since the resulting prior has larger overlap with the likelihood).
Increasing the number of data improves the results, as proves the 43 errors in model selection obtained when $\phi=0.8$ and $D_y=100$. 
\newline
Another way to reduce this problem is to apply the likelihood-based priors (using  the same number of data in the construction of the prior) to both models.
{
This results in using the following intrinsic Bayes factor 
\begin{align}
	\text{IBF}_{12} = \frac{1}{D_y}\sum_{i=1}^{D_y}
	\frac{\int_0^\infty \ell_1({\bf y}_{-i}|\theta)\post_1(\theta|y_i)d\theta}{\int_0^1\ell_2(\y_{-i}|\phi)\post_2(\phi|y_i)d\phi} = 
	\frac{1}{D_y}\sum_{i=1}^{D_y}\frac{\int_0^\infty \ell_1({\bf y}|\theta)d\theta/\int_0^\infty \ell_1( y_{i}|\theta)d\theta}
	{\int_0^1\ell_2({\bf y}|\phi)d\phi
	/\int_0^1 \ell_2( y_{i}|\phi)d\phi}.
\end{align}
We run 100 simulations employing this procedure and observed that the number of errors in detecting the model $\mathcal{M}_2$ when $\phi_\text{true} \in \{0.5, 0.8\}$ gets reduced to, respectively, 18 and  16 when $D_y=30$.
%entonces se reducen bastante los errores que aparecen en la parte derecha de las tablas 3 y 4.... INCLUIR ESTO, AUNQUE SEA DE PALABRA SIN TABLA? 
}

\begin{table*}[!ht]
	\centering
	%\small
	\caption{Model comparison for $D_y=30$. Minimum and maximum $\text{IBF}_{12}$ under true model $\mathcal{M}_1$ (Poisson model) and $\mathcal{M}_2$ (geometric model), over 100 independent runs.}
	\vspace{0.2cm}
	%	{\footnotesize
	\begin{tabular}{|c|c|c|c||c|c|c|c|}
		\hline	
		% & & & & & & &  \\
		\multicolumn{4}{|c||}{True model $=\mathcal{M}_1$} & \multicolumn{4}{|c|}{True model $=\mathcal{M}_2$} \\		  
		\hline
		$\theta$ & min $\text{IBF}_{12}$ & max $\text{IBF}_{12}$ & Errors (IBF$_{12}<1$) &$\phi$ & min $\text{IBF}_{12}$ & max $\text{IBF}_{12}$ & Errors (IBF$_{12}>1$)  \\
		\hline 
		5 & 6.28$\times 10^3$ & 3.95$\times 10^{11}$ & 0 &  0.2 & 1.61$\times 10^{-26}$  & 9.76 & 2  \\ 
		\hline
		2 &  0.55 & 7.40$\times 10^6$ & 1 & 0.5 & 5.45$\times 10^{-9}$  & 884.25 &  30 \\
		\hline
		& & & & 0.8 &  0.004 & 10.51 & 66 \\
		\hline
	\end{tabular}
	%	}	
	\label{table_1}
\end{table*}

\begin{table*}[!ht]
	\centering
	%\small
	\caption{Model comparison for $D_y=100$. Minimum and maximum $\text{IBF}_{12}$ under true model $\mathcal{M}_1$ (Poisson model) and $\mathcal{M}_2$ (geometric model), over 100 independent runs.}
	\vspace{0.2cm}
	%	{\footnotesize
	\begin{tabular}{|c|c|c|c||c|c|c|c|}
		\hline	
		% & & & & & & &  \\
		\multicolumn{4}{|c||}{True model $=\mathcal{M}_1$} & \multicolumn{4}{|c|}{True model $=\mathcal{M}_2$} \\		  
		\hline
		$\theta$ & min $\text{IBF}_{12}$& max $\text{IBF}_{12}$& Errors (IBF$_{12}<1$) &$\phi$ & min $\text{IBF}_{12}$ & max $\text{IBF}_{12}$& Errors (IBF$_{12}>1$) \\
		\hline 
		5 &  2.38$\times 10^{11}$ & 4.52$\times 10^{29}$ & 0 & 0.5 & 1.98$\times 10^{-13}$ & 500.52 & 4  \\ 
		\hline
		2 & 2.22$\times 10^3$  & 2.60$\times 10^{14}$ & 0 & 0.2 & 2.02$\times 10^{-72}$ & 3.34$\times 10^{-18}$ & 0 \\
		\hline
		& & & & 0.8 &  0.003 & 6.69 & 43\\
		\hline
	\end{tabular}
	%	}	
	\label{table_2}
\end{table*}

\subsection{Exoplanet detection}\label{SuperEx}

In recent years, the problem of revealing objects orbiting other stars has acquired large attention. Different techniques have been proposed to discover exo-objects but, nowadays, the radial velocity technique is still the most used \citep{Gregory2011,Barros2016,Affer2019,Trifonov2019}. The problem consists in fitting a dynamical model to data acquired at different moments spanning during long time periods (up to years). The model is highly non-linear and,  for certain sets of parameters, its evaluation is quite costly in terms of computation time.  This is due to the fact that its evaluation involves numerically integrating a differential equation, or using an iterative procedure for solving a non-linear equation (until a certain condition is satisfied). This loop can be very long for some sets of parameters. 

\begin{table}[t]
	\centering
	\caption{Description of parameters in Eq.~\eqref{eq:rv}.}
	\small
	\begin{tabular}{lll} % Column formatting, @{} suppresses leading/trailing space
		\hline
	    \multicolumn{3}{l}{{{\bf For each planet:}}}\\
	    \hline
		Parameter & Description & Units \\
		\hline
		\hline
		$K_i$        & amplitude of the curve & m\,s$^{-1}$ \\
		$\omega_{i}$      & longitude of periastron & rad \\ 
		$e_i$        & orbit's eccentricity    & \ldots \\
		$P_i$        & orbital period        & s \\
		$\tau_i$       & time of periastron passage & s \\
		\hline
		\multicolumn{3}{l}{\footnotesize {{\bf Not depending on the number of planets (below): }}}\\
		\hline
		$V_0$      & mean radial velocity   & m\,s$^{-1}$ \\
		%$\mathbf{s}$         & jitter               & m\,s$^{-1}$ \\
		\hline
		\multicolumn{3}{l}{\footnotesize { {\bf Not inferred directly - it is a function of $e_i$, $P_i$, $\tau_i$ and $t$ (below):}} }\\
		\hline
		${u}_{i,t}$      & true anomaly     & rad \\
		\hline
	\end{tabular}
	\label{tab:rvpar}
\end{table}

%%%----------------------------%%%%

{
	\subsubsection{Model description}
}
%{
%\subsubsection{Model description}
%}
%%%----------------------------%%%%
When analyzing radial velocity data of an exoplanetary system, it is commonly accepted that the wobbling of the star around the centre of mass is caused by the sum of the gravitational force of each planet independently and that they do not interact with each other. 
\newline
Each planet follows a Keplerian orbit and the radial velocity of the host star {(which is our observed noisy measurement $y_t$, at time $t$)} is given by
\begin{align}
{{y}_{t}} &{= f_t({\bm \theta})+\xi_t,} \nonumber \\ 
{y}_{t} &= V_0 + \sum\limits_{i = 1}^{S} K_i \left[ \cos \left( {u}_{i,t} + \omega_{i} \right) + e_i \cos \left( \omega_{i} \right) \right] +\xi_t,
\label{eq:rv}
\end{align}
with $t=1,\ldots,T$,\footnote{More generally, we can have $y_{t_j}$ with $j=1,...,T$.} { where  $\xi_t$ is a Gaussian noise perturbation with variance $\sigma_e^2$, 
 $S$ is the number of planets and the variable of interest $\x$ is the vector of dimension $D_{\x}=1+5S$,
\begin{align*}
\x= [V_0, K_1, \omega_{1}, e_1, P_1, \tau_1, \ldots, K_S, \omega_{S}, e_S, P_S, \tau_S].
\end{align*}
The meaning of each parameter is given in Table~\ref{tab:rvpar}. We have set 
\begin{align}
f_t({\bm \theta})=V_0 + \sum\limits_{i = 1}^{S} K_i \left[ \cos \left( {u}_{i,t} + \omega_{i} \right) + e_i \cos \left( \omega_{i} \right) \right].
\end{align} 
We observe the vector ${\bf y}=[y_1,...,y_T]$ of noisy measurements. The so-called {\it true anomaly} ${u}_{i,t}$ is function of $t$, $e_i$, $P_i$ and $\tau_i$, as we described in the next subsection. It represents the angular position of the $i$-th exoplanet in its orbit with respect to the periastron. The assumption of no correlation in the noise is settled in the nature of the data. The radial velocity is an indirect measure that is determined through the combination of thousand of individual measures at each observation. }

{
\subsubsection{Computation of $u_{i,t}$ and evaluation of the nonlinearity $f_t$} 
%}

%Its dependence on time is given by 
%%
%\begin{equation}
%	\frac{du_i}{dt} = \frac{2\pi \left( 1+e_i \cos u_i \right)^2}{P_i(1 - e_i^2)^{3/2}}.
%\end{equation}
%The solution of the differential equation can be obtained from the Kepler's laws. 

The true anomaly $u_{i,t}$ is related to  $e_i$, $P_i$ and $\tau_i$, by the following equations:
\begin{align}
%	M_{i,t} & = & , \\[3mm]
 &u_{i,t} =   2 \arctan \left(\sqrt{\frac{1 + e_i}{1 - e_i}} \tan \frac{E_{i,t}}{2}\right), \label{eq:trueAnomaly} \\
	& E_{i,t} - e_i \sin \left( E_{i,t} \right)=\frac{2\pi}{P_i} \left( t - \tau_i \right).  \label{iterativeEq}
	\end{align}
	Hence, we need to solve the Eq. in \eqref{iterativeEq} in order to obtain the value $ E_{i,t} $ and then replace in Eq. \eqref{eq:trueAnomaly}. The solution to Eq. \ref{iterativeEq} is found iteratively applying a Newton-Raphson procedure \citep{martino2021automatic,lopez2021bayesian}.} For certain sets of parameters, this iterative procedure can be particularly slow and the computation of the likelihood becomes quite costly. 
	\newline
	{ As an example, let us set $S=1$ for the sake of simplicity. Given a value of $\x^*=[V_0^*, K_1^*, \omega_{1}^*, e_1^*, P_1^*, \tau_1^*]$, in order to evaluate $f_t(\x^*)$ we proceed as follows:
\begin{enumerate}
\item  Given $e_1^*$, $P_1^*$, and $\tau_1^*$, compute approximately the values of $E_{1,t}$'s for each $t$,  from  Eq. \eqref{iterativeEq}, by applying the Newton-Raphson method. 
\item  Given the values  $E_{1,t}$'s previously obtained,  compute $u_{1,t}^*$ for each $t$.
\item  Given the values $u_{1,t}^*$'s previously obtained, and $V_0^*$, $e_1^*$, $\omega_1^*$,  compute $f_t(\x^*)$ for each $t$.
\end{enumerate}
A periodic link between the variables $\tau_1^*$ and $\omega_1^*$ could appear and, as a consequence, the likelihood function could have multiple equivalent (periodic) modes. This link can be broken by a proper choice of the priors.
%As final remark, 	
}	

{
	\subsubsection{Likelihood function and model evidence}
	}
 For a single object (e.g., a planet or a natural satellite), the dimension of $\x$ is $D_{\x} = 5+1=6$, with two objects the dimension of $\x$  is $D_{\x} = 11$, etc. 
The Eq. \eqref{eq:rv} induces a likelihood function, i.e., 
$$
\ell({\bf y}|\x,\sigma_e)=\prod_{t=1}^T\ell(y_{t}|\x,\sigma_e),
$$
%\end{align*}
where ${\bf y}=\{y_{1},\ldots,y_{T}\}$.
%and $\ell(y_t|\x,\sigma_e) = \mathcal{N}(y_t|V_0 + \sum\limits_{i = 1}^{S} K_i \left[ \cos \left( {u}_{i,t} + \omega_{i} \right) + e_i \cos \left( \omega_{i} \right) \right],\sigma_e^2) $. 
%Given a prior density $g(\x)$, the marginal likelihood is given by
%$$
%Z=p({\bf y}|\sigma_e)=\int_{\bm \Theta} \ell({\bf y}|\x,\sigma_e) g(\x) d\x.
%$$ 
Our goal is to infer the number $S$ of planets in the system. 
For this purpose, given prior densities $g_i(\x_i)$ for each model, we have to approximate the model evidences, 
$$
Z_i = \int_{{\bm \Theta}_i} \ell(\y|\x_i,\sigma_e)g_i(\x_i)d\x_i.
$$ 
For simplicity, we consider the noise variance $\sigma_e^2$ is given.

%%%----------------------------%%%%
\subsubsection{Experiments}
\begin{figure}[h!]
	\centering
	\centerline{
	\subfigure[]{\includegraphics[width=0.5\textwidth]{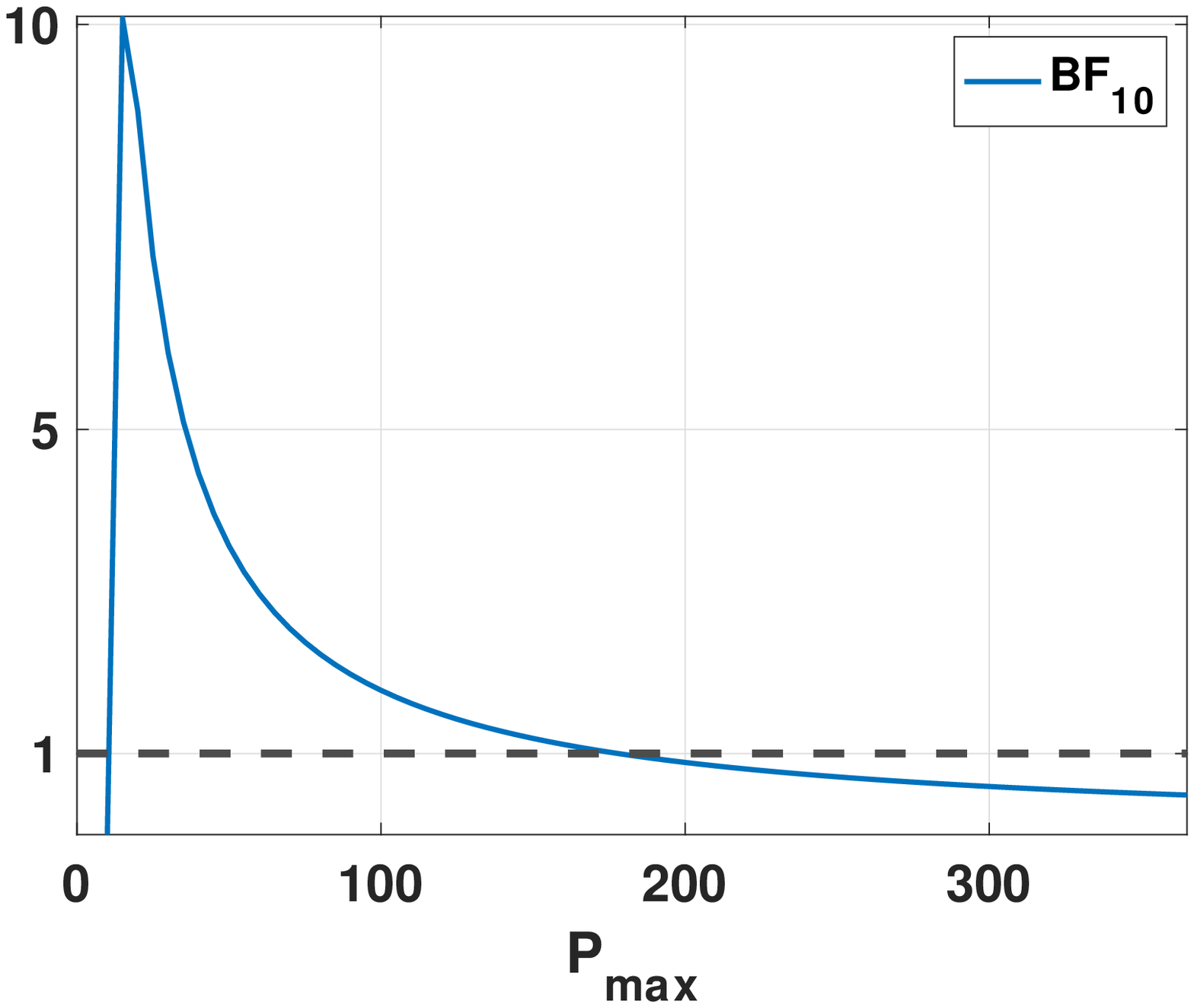}}
	\subfigure[]{\includegraphics[width=0.5\textwidth]{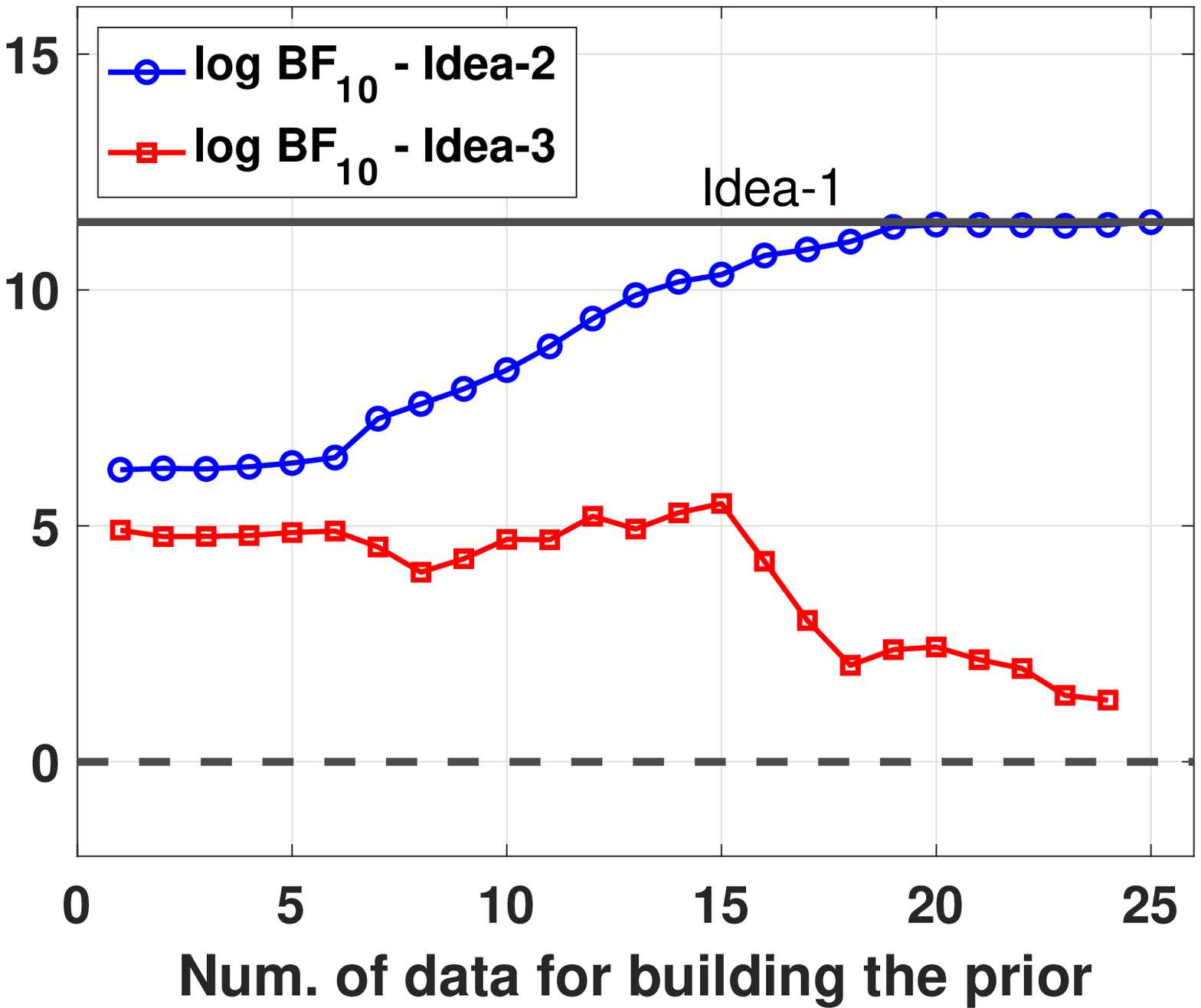}}
	}
	\caption{\label{fig_Pmax}	
	\small {\bf (a)} The Bayes factor BF$_{10}$ as a function of prior width $P_\text{max}$. 
	 Increasing $P_\text{max}$ (i.e. making the prior for $P_1$ more diffuse) eventually produces BF$_{10}<1$.
	 Note also that, when $P_\text{max}$ is small (lower than $P_\text{max}=15$), we have BF$_{10}<1$, preferring the model with zero planet $\mathcal{M}_0$.  { {\bf (b)}   
	 The log-BF$_{10}$ obtained using likelihood-based priors in both models (specifically, the ideas 2 and 3 in Section \ref{aquiVALUCALOCO}) adding sequentially data in the prior construction. Note that  log-BF$_{10}>0$ preferring always (and correctly) the model with one planet. Figure (b) also shows Idea-1 as limit of Idea-2, which provides an upper-bound for the rest of values.     } 
	}
\end{figure}

%%%----------------------------%%%%
Let us denote $\mathcal{M}_0$ and $\mathcal{M}_1$ the models corresponding to zero and one planets.
We generate a set of data $\y$ according to the model with one planet and parameter values $V^\text{true}_0=5$, $K^\text{true}_1=25$, $\omega^\text{true}_1=0.61$, $e^\text{true}_1=0.1$, $P^\text{true}_1=15$, and $\tau^\text{true}_1=3$.
We consider $D_y=25$ total number of observations. All the data are generated with $\sigma_e^2=15$.
The rest of trajectories are generated according to the transition model (and the corresponding measurements $y_{t}$ according to the observation model). 
Our goal is to compute the ratio 
$
\mbox{BF}_{10}=\frac{Z_1}{Z_0},
$
 where $Z_1$ and $Z_0$ denote respectively the marginal likelihood of the model with zero planet and the model with one planet.
As we commented above, the model with zero planet has only one parameter, namely, $\x_0=V_0$ and we choose a uniform prior $\mathcal{U}([-20,20])$. 
For simplicity, in the model with one planet we consider only two degrees of freedom, i.e., $\x_1=[V_0,P_1]$. The rest of parameters are set to their true values. 
We use the same prior for $V_0$ in $\mathcal{M}_1$.
For the period $P_1$, we use  $\mathcal{U}([0,P_\text{max}])$ with $P_\text{max}>0$. 
Namely, we use a uniform prior with varying width. When $P_\text{max}=365$, we are considering a uniform prior over all the possible values of $P$. 
We know that BF$_{10}$ should be greater than $1$ since the data were generated according to model 1.
However, we aim to show that increasing $P_\text{max}$ (which corresponds to use a prior that is more diffuse) makes that BF$_{10}$ eventually becomes smaller than 1. %{(again, this is the Lindley-Bartlett paradox)}.
For the computation of $Z_0$ and $Z_1$ we use a very thin grid within the prior bounds.
In Figure  \ref{fig_Pmax}(a), we show the Bayes factor as a function of $P_\text{max}$. For $P_\text{max}$ greater than 200, we have BF$_{10}<1$, that is, we wrongly choose the model with zero planets. This illustrates again the problematic with the use of vague priors. 
%{
%Indeed, it shows that the uniform prior over the domain of $P_1$ (namely, $[0,365]$) may produce problems in model selection. 
%}

\noindent
{\bf Hierarchical solution.}
Let us denote as $Z_1(P_\text{max})$ the marginal likelihood {of model $\mathcal{M}_1$} for each given value of $P_\text{max}.$
We consider the extended posterior where we use a {hyperprior} for $P_\text{max}$, $g_h(P_\text{max}) = \mathcal{U}([10,365])$, hence the new marginal likelihood is 
$$
Z_{\texttt{new},1} = \int_{10}^{365} Z_1(P_\text{max})g_h(P_\text{max})dP_\text{max}.
$$
The value of $Z_{\texttt{new},1} $  is $9.1095\times 10^{-44}$, which is greater than $Z_0=5.4601\times 10^{-44}$. Hence, with this hierarchical modeling, we select the true model.  
{Note that $g_h(P_\text{max}) = \mathcal{U}([10,365])$ is virtually the more diffuse{d} hyper-prior that we can use in this experiment, since the parameter $P_1$ represents a period of rotation (measure{d} in ``days''), so it varies between 0 and 365. }
\newline
{
{\bf Likelihood-based priors.} Another possible solution is to employ  likelihood-based priors. We apply the Idea-2 and Idea-3 given in Section \ref{aquiVALUCALOCO} to both models. In Idea-2,  a subset of data is used twice (for building the prior and in the likelihood as well) whereas, in Idea-3, the data are split in training (for building the prior) and test (used only in the likelihood). Note that, if we use all the data ($D_y=25$) for building the prior, Idea-2 becomes Idea-1 in  Section \ref{aquiVALUCALOCO}.  
 We start building the prior with only one datum (the first one), and compute the corresponding BF$_{10}$. Then, we add sequentially the rest of data, starting from the second one, until we consider the $25$-th data for Idea-2, and the $24$-th data for Idea-3. The log-BF$_{10}$ is given in Figure \ref{fig_Pmax}(b). In this case, we always choose the true model.  As expected, Idea-2 tends to {favour} the more complex model with respect to Idea-3. Again as expected, Idea-1 provides an upper bound for the BF$_{10}$ obtained by  Idea-2 and Idea-3.
}

\section{Conclusions}

\noindent In this work, we have highlighted some important considerations regarding the computation of marginal likelihoods, which are fundamental quantities for Bayesian model selection. We have discussed the dependence on the choice of the prior density and shown some comforting asymptotic results. Moreover, we have remarked that the use of  improper priors is not suitable for model selection. More generally, we have also discussed {that} the use of diffuse priors, whether proper (vague priors) or improper, are actually very informative for the  model selection procedure (Level-2 of inference).
We have shown by means of illustrative examples the potential pitfalls of using vague priors, and we have provided and discussed several possible solutions for these scenarios, such as the construction of likelihood-based or model-based priors, and partial/fractional Bayes factors. We have also described an alternative for Bayesian model selection to the marginal likelihood approach, called posterior predictive. Furthermore, the connection with the information criteria has been also presented.
%and checked the usefulness of the solutions proposed for improper priors.
One of the considered numerical experiment is a real-world astronomical application, consisted on detecting the number of objects orbiting a star. 
\newline
{
We list below some final highlights of the work:
\begin{itemize}
\item Clearly, for a finite number of data $D_y$, the results of Bayesian inference depends on the choice of the prior densities. However, the Bayesian model selection (based on the model evidence $Z$) is consistent, i.e., selects the true/best model as $D_y \to \infty$, under very mild assumptions on the prior densities.
\item Improper priors are not allowed in Level-2 since the marginal likelihoods are undetermined.
\item Considering a a finite number of data $D_y$, uniform priors can be highly informative in model selection, i.e., the Level-2 of inference (unlike in Level-1). 

\item As a consequence of the previous points, in absence of a-priori information, there is a need of procedures for designing {\it objective} priors for the Level-2 of inference. The construction of objective priors is generally based on data, likelihood functions and/or observation models. The simplest scheme,  in this sense, is the {\it empirical Bayes} approach, where the prior parameters are tuned maximizing the marginal likelihood. Other more sophisticated schemes use parts of the data for building a suitable objective prior. 
 %We enumerate below the main ideas in increasing order of complexity: 
%\begin{itemize}
%\item  
%\item 
%\item More generally, in a full-Bayesian approach of a  {\it hierarchical modeling}, we compute an averaged marginal likelihood averaged taking into account  {\it several} prior densities, all the members of the same parametric family. Clearly, the prior densities in the ....   
%\end{itemize}

\item Alternative approaches to standard Bayesian model selection (which is based on the model evidence $Z$) rely on the concept of {\it prediction} (recalling the frequentist idea of cross-validation). These approaches seems to be more robust with respect to the choice of the prior densities, but the consistency is not generally ensured \citep{vehtari2012survey}.

 %(i.e., selecting the true model as $D_y \to \infty$)
\end{itemize}

} 

%%%%A connection with information criteria is briefly presented.

{\footnotesize
	\section*{Acknowledgments}
	The authors would like to thank the two anonymous referees for their detailed comments and suggestions. 
	This work has been supported by Spanish government via grant FPU19/00815 and by Agencia Estatal de Investigaci\'on AEI (project SP-GRAPH, ref. num. PID2019-105032GB-I00).
	
}

\bibliographystyle{apacite}
\bibliography{bibliografia,bibliografia_otro,bibliografia_otro2}

\begin{appendices}

{
%%%%%%%%%%%%%%%%%%%%%%%%%%%%%%%%%%%%%%%%%%%
\section{Implicit model penalization contained in $Z$}\label{LucaImplicitPEN}
%%%%%%%%%%%%%%%%%%%%%%%%%%%%%%%%%%%%%%%%%%%

The illustrative example in Section \ref{LucaParaSection} allows us to show that the marginal likelihood $Z$ contains an implicit model penalization  \citep[Ch. 28]{mackay2003information}. In that example, we consider the uniform prior $g(\x) = \frac{1}{|B|} \bm{1}_{B}(\x)$, where $|B|$ represents the volume of $B$. Without loss of generality, let us consider the case of $B$ being a hypercube centered at the origin with side length $\delta$, i.e, $B = \left[-\delta/2,\delta/2\right]^{D_{\x}} \subseteq {\bm \Theta}$,
with  volume  $|B| = \delta^{D_{\x}}$. 
 From Eq. \eqref{SuperImpZlucaParadox}, we have 
 	\begin{align}
		\log Z &= \log\int_{B} \ell(\y|\x)d\x  - \log |B|, \nonumber \\
		&= \log\int_{B} \ell(\y|\x)d\x  - D_{\x}\log \delta.
		%	\\
		%	&= \log \ell_\text{max} + \log \frac{Z_\ell(B)}{\ell_\text{max}}  - d_{\x}\log \delta, 
	\end{align}
Note that both terms depend on the size $\delta$ and the dimensionality $D_{\x}$.\footnote{$B$ depends on  both $\delta$  and $D_{\x}$, whereas the $\ell(\y|\x)$ depends on  $D_{\x}$.}
For a fixed $D_{\x}$, increasing $\delta$ affects both the fitting and penalty terms. Both terms grows as $\delta$ increases. However, note that while the first term is bounded by $S_{D_{\x}}= \int_{{\bm \Theta}} \ell(\y|\x)d\x$,\footnote{ $B$ and ${\bm \Theta}$ depend both on the parameter dimension $D_{\x}$. Hence, also $S$ depends on $D_{\x}$.  For this reason, here we use the more proper notation $S_{D_{\x}}= \int_{{\bm \Theta}} \ell(\y|\x)d\x$.} 
and the second term can grow indefinitely in $\delta$. Hence, we have the following upper bound for $\log Z$, that is 
\begin{align}
 \log Z \leq\underbrace{\log S_{D_{\x}}}_{\texttt{fitting}} \underbrace{- D_{\x}\log \delta}_{\texttt{penalty}},
\end{align}
where we can interpret the first term in the above equation as a fitting term, and the second term as a penalty term over the model complexity/order \citep[Ch. 28]{mackay2003information}.

{\rem This penalty term can also be interpreted as an {\it implicit log-prior} term over the corresponding model. }
\newline
\newline
Moreover, for $ \delta \to \infty$, we have $\log Z \to -\infty$ (keeping fixed $D_{\x}$). Similar considerations and the connection with information criteria are also given in the following Appendix \ref{ObradeArteCriteria}. }

\section{Marginal likelihood $Z$ and information criteria}\label{ObradeArteCriteria}

The marginal likelihood can be expressed as
\begin{align}\label{SuperIMPEqZ}
Z= \ell_\text{max}W,
\end{align}
where  $W \in[0,1]$ is the {\it Occam factor}  \citep[Sect. 3]{knuth2015bayesian}.  More specifically, the Occam factor is defined as 
\begin{align}
W = \frac{1}{\ell_\text{max}}\int_{\bm \Theta} g(\x) \ell(\y|\x)d\x,
\end{align}
and it is $\frac{\ell_\text{min}}{\ell_\text{max}}\leq W \leq 1$. The factor $W$ measures  the penalty of the model complexity {\it intrinsically} contained in the marginal likelihood $Z$: this penalization depends on the chosen prior and the number of data involved.
\newline
\newline
 Considering the expression \eqref{SuperIMPEqZ} and taking the logarithm, we obtain
\begin{align}\label{Zlogaqui}
\log Z= \log\ell_{\max}+ \log W
%&=\log \ell_{\max}+\log \Delta_\ell-D_\theta \log \Delta_\theta,  \nonumber 
%\\
%&=\log \ell_{\max}+\eta D_\theta,  \label{aquiEqDtheta}
\end{align}
%where $\eta=\frac{\log \Delta_\ell}{D_\theta}-\log \Delta_\theta$ is a constant value, which also depends on the number of data $D_y$ and, generally, $\eta=\eta(D_y,D_\theta)$. 
%Different model selection rules in the literature consider the simplification $\eta=\eta(D_y)$. 
Note that $\log \ell_{\max}$ is a fitting term whereas $\log W$ is a penalty for the model complexity. %%%%{ Note that }
Instead of maximizing $Z$ (or $\log Z$) for model selection purposes, several authors consider the {\it minimization} of some cost functions $C$ derived by different information criteria \citep{schwarz1978estimating,Hannan79,Spiegelhalter02}. Most of the criteria, suggested in the literature, can be expressed as 
%To connect them with the marginal likelihood maximization, we consider the expression of $-2 \log Z=-2 I$ where $I=-\log Z$ resembles the Shannon information associated to $Z=p({\bf y})$ , i.e.,
%To connect to BIC and AIC (we want to minimize them):
\begin{eqnarray}\label{Caqui}
C=\underbrace{-2\log \ell_{\max}}_{\mbox{\scriptsize fitting }}\underbrace{+ 2 \eta D_{\x}}_{\mbox{\scriptsize penalization}},
\end{eqnarray}
%-2\log Z
where $\eta$ is a real value that is often chosen as function of the number of data $D_y$, and $D_{\x}$ is the dimension of $\x$, i.e., the number of parameters. The first term is a fitting term (which fosters the choice of more complex models), whereas the second one is a model penalization term (which promotes the choice of simpler models).
\begin{rem}
Note that the expression of $C$ is similar to 
$$
-2 \log Z=-2\log\ell_{\max}-2 \log W,
$$
considering  Eq. \eqref{Zlogaqui}, where $-2 \log W$ plays the role of the second factor $2 \eta D_{\x}$ in Eq. \eqref{Caqui}.
\end{rem}
\noindent
The expression \eqref{Caqui} encompasses several well-known information criteria proposed in the literature and shown in Table \ref{TablaIC}, which differ for the choice of $\eta$. %In all these cases, $\eta$ is just a function of the number of data $D_y$.  %{ However, since $\eta$ depends on $\Delta_\ell$, it should depends also on $D_\theta$....}
%More details regarding these information criteria are given in Section \ref{ExploitingFunctionalIdentity}. 

{\rem The penalty term $2 \eta D_{\x}$ in the information criteria is the same for every parameter. The Bayesian approach allows the choice of different penalties, assuming different priors, one for each parameter, i.e., for each component of $\x$. }

\begin{table}[!h]	
	%{
	\caption{Different information criterion for model selection.}\label{TablaIC}
	\vspace{-0.3cm}
	\footnotesize
	\begin{center}
		\begin{tabular}{|c|c|} 
			\hline 
			{\bf Criterion} & {\bf Choice of } $\eta$   \\ 
			\hline 
			\hline 
			Bayesian-Schwarz information criterion (BIC) \citep{schwarz1978estimating} &  $\frac{1}{2}\log D_y$ \\
			\hline 
			Akaike information criterion  (AIC) \citep{Spiegelhalter02} &  $1$ \\
			\hline 
			Hannan-Quinn information criterion (HQIC)  \citep{Hannan79}&  $\log(\log(D_y))$ \\
			\hline 
		\end{tabular}
	\end{center}
	%	}
\end{table}

\end{appendices}

\end{document}